\documentclass[pdftex,epsf]{iopart}

\usepackage{tikz,tikz-3dplot,pgfplots}
\usepackage{amsfonts,dsfont}
\usepackage{amssymb}
\usepackage{amsthm}
\usepackage{bm} 
\usepackage{xcolor}

\usepackage{float}
\usepackage[utf8]{inputenc}
\usepackage[T1]{fontenc}
\usepackage[english]{babel}
\usepackage{lmodern}
\usepackage{showlabels,soul}

\usepackage{graphicx}
\usepackage[hidelinks]{hyperref}
\usepackage[english]{babel}

\newlength{\bredde}
\def\slash#1{\settowidth{\bredde}{$#1$}\ifmmode\,\raisebox{.15ex}{/}
\hspace*{-\bredde} #1\else$\,\raisebox{.15ex}{/}\hspace*{-\bredde} #1$\fi}
\newcommand{\Nf}{N_{\rm f}}
\newcommand{\tNf}{\widetilde{N_{\rm f}}}
\newcommand{\ulbackslash}[2]{{\raisebox{-.2em}{$#1$}\left\backslash\raisebox{.2em}{$#2$}\right.}}
\newcommand{\sign}{\,{\rm sgn}\,}

\begin{document}

\title[Level spacings in the chiral Gaussian unitary ensemble]{Consecutive 
level spacings in the chiral Gaussian unitary ensemble:\\ From the hard and soft edge to the bulk}
\author{{\sc G. Akemann$^1$, V. Gorski$^1$ and M. Kieburg$^2$}\\~\\
$^1$Faculty of Physics,
Bielefeld University,\\
Postfach 100131,
D-33501 Bielefeld, Germany\\
$^2$
School of Mathematics and Statistics, University of Melbourne,\\ 813 Swanston Street, Parkville, Melbourne VIC 3010, Australia}

\begin{abstract}
The  local spectral statistics of random matrices 
forms 
distinct universality classes,  strongly depending on the position in the spectrum.
Surprisingly, the spacing between consecutive eigenvalues at the spectral edges has received little attention,
where the density diverges or vanishes,  respectively. This different behaviour is called hard or soft edge.
We show that 
the spacings at the edges 
are almost indistinguishable from the spacing in the bulk
of the spectrum. We present analytical  results for consecutive spacings between the $k$th and $(k+1)$st smallest eigenvalues in the chiral Gaussian unitary ensemble, both for finite- and large-$n$. The result depends on the number of the generic zero modes $\nu$ and the number of flavours $N_{\rm f}$, which are given in terms of characteristic polynomials, as motivated 
by Quantum Chromodynamics (QCD). We find that the convergence in $n$ is very rapid. The same can be said separately about the limit $k\to\infty$ (limit to the bulk) and $\nu\to\infty$ (limit to the soft edge). Interestingly, the Wigner surmise is 
a 
very good approximation for all these cases and, apart from $k=1$,  shows a deviation below one percent. These findings are  corroborated with Monte-Carlo simulations.
We finally compare for $k=1$ with data from QCD on the lattice, being in this symmetry class.

\end{abstract}

\maketitle

\section{Introduction}\label{sec:intro}

One of the cornerstones in 
applying random matrix theory is the local level spacing distribution. It provides a statistical description of the spacing between consecutive eigenvalues. For that purpose,  the mean level density around the point where we zoom in has to be unfolded, meaning it is mapped to be approximately constant, see~\cite{GMW}. Then, the energies are measured in units of this constant. This procedure applies to the bulk of the spectrum. 
The spacing distribution continues to be a popular tool in today's applications, ranging from quantum  spin chains \cite{SPT}, quantum circuits \cite{KOR}, chemical isotopes 
\cite{AM1} and graphene \cite{AM2} to classical integrable systems \cite{EG}
and 
theoretical studies \cite{RDC}, which are all from past year.
We refer to~\cite{GMW,Handbook} for a list of more topics and references therein.

Let us also look back at the classical papers where
the spacing distribution has 
been proposed 
in the study of quantum systems displaying chaotic behaviour, 
the Bohigas-Giannoni-Schmit conjecture~\cite{BGS}, cf. \cite{CGV}: quantum systems that are fully chaotic follow random matrix statistics. This field of research has been subsumed and influenced  by Fritz Haake's famous book \cite{Haake} in several editions. 
Let us mention an important mathematical contribution of Fritz Haake in this context. Together with Barbara Dietz \cite{DH}, he has carried out a precise expansion of the true spacing distribution for the classical random matrix ensembles for comparison with data, that we will use as well. This expansion is the one developed by Pad\'e~\cite{Pade}, who introduced an expansion in rational functions instead of polynomials, as it is common for the Taylor expansion.  This approximation for the level spacing distribution by Dietz and Haake  goes beyond the frequently used Wigner surmise. At the same time, it proves a practical expression compared to a truncation of the infinite product of Fredholm determinant eigenvalues in Mehta's book~\cite{Mehta}, given in terms of an integral over spheroidal functions.
In parallel, Fritz Haake has pursued an impressive research programme to prove random matrix statistics from a semi-classical expansion, cumulating in~\cite{NJPFH}. This has lead to a much deeper understanding of chaotic quantum systems.

A second example where the random matrix approximation is very well understood is the low-energy Dirac operator spectrum in Quantum Chromodynamics (QCD) introduced in~\cite{SV93}, which has motivated our study. Here, the spectral edge 
representing a hard edge  is focussed on, zooming into the origin where the lowest eigenvalues are located. Without going into detail of the vast literature on the application to QCD, and 
on comparing 
to lattice data from first principle lattice QCD simulations, cf. \cite{TJ,reviewQCD}, let us focus on random matrix questions. In this field, typically the local density correlation functions, given by the universal Bessel-kernel~\cite{ADMN}, and the distributions of the smallest eigenvalues~\cite{DNW,GWW,DN} are employed for comparison. However, few works have applied the level spacing distribution as a measure for random matrix statistics, both at the hard edge~\cite{Kim,Poul,Glozman} and in the bulk of the spectrum~\cite{Thomas}. 
Averaging over several consecutive spacings, a good agreement was found at the hard edge below the chiral phase transition. Only in the vicinity of the phase transition deviations from the bulk spacing were found~\cite{Kim,Poul,NGTKP,KovacsPittler}. 
It came as a surprise to us, that no analytical results were available at the hard edge for the spacing distribution, and that at least on average the bulk spacing leads to a good description. 

In the present article, we will provide a closed form analytical  expression for the spacing distribution in the chiral Gaussian unitary ensemble ($\chi$GUE) that is relevant for QCD with three colours in the fundamental representation, and other strongly interacting field theories which lie in the same universality class.
We consider the spacing between two consecutive eigenvalues close to the origin, beginning with the smallest and second smallest eigenvalue. Our derivation follows ideas from  the computation of the distribution of individual eigenvalues at the hard edge, starting with the smallest one~\cite{DNW,GWW,DN}. The resulting spacings depend on several parameters of the ensemble, the number of exact zero eigenvalues $\nu$, that is related to the gauge field topology, and a fixed number of (light) quark flavours $N_{\rm f}$ with masses $m_f$. Our results hold for finite matrix size $n$ and in the asymptotic microscopic scaling limit $n\to\infty$. It provides an alternative to the somewhat heavy machinery of analysing the Fredholm determinant of the underlying integral kernel, that typically leads to an expression including Painlev\'e V differential equations~\cite{TWBessel}. In contrast, we will obtain expressions in terms of $k$-fold integrals over a determinant of Bessel-functions of size $N_{\rm f}+\nu+k$ 
for the $k$th spacing.

Moreover, we will find that, both at finite (even small) matrix dimension $n$, as well as asymptotically at large $n$, the spacings at the hard edge are very close to the GUE bulk spacing distribution, including a very weak dependence on $\nu$ and $N_{\rm f}$. The deviations range from one percent to only a few per mill when varying the parameters.  Hence a comparison to QCD lattice data, that we also undertake, does not allow to discriminate between the two, even for the smallest to second smallest eigenvalue spacing 
at $k=1$. The situation at the soft edge is very similar, where the random matrix eigenvalue density vanishes as a square root. Here, apparently so far only the local Airy-kernel~\cite{Peter} and distribution of the largest eigenvalue~\cite{TWAiry} were studied, but not the spacing distribution.
Even if we cannot offer analytical  results in the soft-edge case, our numerical investigation suggests that also the spacing between the largest and second largest eigenvalue is very close to the bulk spacing. This is numerically corroborated at the inner soft edge when $\nu\gg1$, too.

Let us emphasise that we are studying the transition between two different local random matrix statistics, that is in our case the Bessel-kernel and sine-kernel statistics. Those statistics also appear in the transition ensembles between the $\chi$GUE and GUE, where the former ensemble has chiral symmetry. Such a transition has been studied on the level of correlation functions \cite{AN,MK}, where it is very pronounced and relevant for the Wilson Dirac operator spectrum \cite{DSV}. In contrast, for the observable of consecutive level spacings the difference between the two ensembles turns out to be tiny at the origin (the hard edge), and we are almost immediately in the bulk. 
The transition between different random matrix symmetry classes is a classical question and has been studied in the bulk \cite{MehtaPandey} and at the edges \cite{FNH}. We will not touch upon another transition, between quantum chaotic and integrable behaviour which is generically Poisson, compare \cite{Haake}.

The remaining article is organised as follows. In Section~\ref{sec:heuristic}, we study the limit from the hard and soft edge to the bulk, by taking the large argument asymptotic of the universal Bessel- and Airy-kernel, respectively, that yields the universal sine-kernel in the bulk. This provides a convergence rate at the level of the density correlation functions. Section~\ref{sec:gap} discusses general expressions for the gap probability and 
$k$th 
spacing distributions in terms of a $k$-fold integral over the ratio of partition functions with $N_{\rm f}+\nu+2k$ flavours. These are then evaluated in Section~\ref{sec:finiteN} for the $\chi$GUE at finite matrix dimension $n$, yielding compact determinantal expressions of the size of $N_{\rm f}+\nu+k$ flavours. In Section~\ref{sec:largeN}, we take the large-$n$ limit at fixed $\nu$ and $N_{\rm f}$ which is known to be universal. The convergence of our analytical  results for the hard edge spacing towards the bulk spacing is illustrated for quenched ($N_{\rm f}=0$) and unquenched settings ($N_{\rm f}\neq0$), and compared to the approximation of the bulk spacing through the Wigner surmise.  
Section~\ref{subsec:LQCD_data} contains the comparison to data from lattice QCD, and in Section~\ref{sec:conclusio} we present our conclusions.  In \ref{spaceDH} we collect the exact expression for the GUE bulk spacing distribution from~\cite{DH}.

\section{Transition from Bessel- and Airy- to the Sine-Kernel}\label{sec:heuristic}

The spacing between consecutive eigenvalues of random matrices with unitary symmetry in  
the bulk of the spectrum is well-approximated by Wigner's surmise~\cite{Haake,GMW} for $2\times2$ GUE matrices,
\begin{equation}\label{Wignersurmise}
p_{\rm Wigner}(s)=\frac{32s^2}{\pi^2}e^{-4s^2/\pi}.
\end{equation}
We will also give a 
much more precise 
expression in \ref{spaceDH}, cf. \cite{Mehta}, which is universal.
In the present case of the $\chi$GUE also called complex Wishart-Laguerre ensemble, 
deviations should show up at the hard edge because 
there the $k$-point correlation functions differ from the ones in the bulk of the spectrum.

Before we address this question in terms of the nearest neighbour level spacing distribution in detail, see Section~\ref{sec:gap}, let us investigate the transition on the level of the kernel of the underlying limiting determinantal point process, that is the transition from the Bessel-kernel at the hard edge to the sine-kernel in the bulk. 
Additionally, we also discuss the transition from the Airy-kernel at the soft edge to the sine-kernel 
in the bulk 
for completeness. Although we do not have analytical  results at the soft edge, we will numerically study this transition for the spacing as well in Section~\ref{sec:asymp.nu}.

The eigenvalue statistics of the $\chi$GUE follows a determinantal point process. This means the $k$-point correlation function of the eigenvalues has the form
\begin{equation}\label{k-point-det}
R_k(x_1,\ldots,x_k)=\det[K(x_a,x_b)]_{a,b=1,\ldots,k},
\end{equation}
with correlation kernel $K(x_a,x_b)$. This holds already at finite matrix size, see~\eref{k-point-det.n} in Section~\ref{sec:finiteN}, as well as in the asymptotic scaling limit in the bulk, at the hard and soft edge. For example, the microscopic level density is given by $\rho(x)=R_1(x)$.

The  difference between the  eigenvalue statistics at different locations in the spectrum can be seen in the kernel which is the sine-kernel~\cite{Mehta} inside the bulk of the spectrum,
\begin{equation}\label{sine-kernel}
K_{\rm sine}(x_a,x_b)=\frac{\sin[\pi(x_a-x_b)]}{\pi(x_a-x_b)},
\end{equation}
and the Bessel-kernel~\cite{Peter,TWBessel} at the hard edge of the spectrum,
\begin{equation}\label{Bessel-kernel}
\hspace*{-1cm} K_{\nu}(x_a,x_b)=\sqrt{x_ax_b}\frac{x_b J_\nu( x_a)J_{\nu-1}( x_b)-x_a J_\nu( x_b)J_{\nu-1}( x_a) }{x_a^2-x_b^2}, 
\end{equation}
where $J_\nu$ is the Bessel function of the first kind and the index $\nu$ counts the number of zero eigenvalues, measuring the strength of the repulsion from the origin. The Airy-kernel~\cite{Peter,TWAiry}
\begin{equation}\label{Airy-kernel}
K_{\rm Airy}(x_a,x_b)=\frac{{\rm Ai}(x_a){\rm Ai}'(x_b)-{\rm Ai}'(x_a){\rm Ai}(x_b)}{x_a-x_b}, 
\end{equation}
is obtained at the soft edge of the spectrum, with ${\rm Ai}(x)$ and ${\rm Ai}'(x)$ the Airy function (of the first kind) and its first derivative. All three kernels are universal for a broad class of systems and hold for a much larger class of 
unitarily invariant 
ensembles of random matrices 
than Gaussian, see the review by  Kuijlaars~\cite[Chapt. 6]{Handbook}. 
For an extension to orthogonal and symplectic ensembles see the book by Deift and Gioev \cite{DG}.

In principle, all spectral information is contained in the kernels in the respective region of the spectrum, including the spacing distribution. To assess the latter, one typically uses the gap probability, the probability $\mathbb{P}_{[a,b]}$ that an interval $[a,b]$ does not contain eigenvalues, expressed in terms of the Fredholm determinant of the corresponding kernel,
\begin{equation}
\mathbb{P}_{[a,b]}=\det[1-K_{[a,b]}],
\end{equation}
where the integral operator is
\begin{equation}
K_{[a,b]}\phi(x)=\int_a^bK(x,y)\phi(y) dy,
\end{equation}
for an arbitrary ($L^2$-integrable) test function $\phi$.
 The spacing distribution then follows by differentiation, as it is explained in more detail in Section~\ref{sec:gap}. However, we will not follow this route of Fredholm determinants and compute the spacing distribution in a different manner. 

The second information that can be retrieved from the kernel is the mean level spacing. It follows from the microscopic level density at large argument, which yields the macroscopic level density at that point in the spectrum, which gives the inverse mean spacing. This quantity is important for the unfolding of the spacing distribution. 
From the sine-kernel we have 
\begin{equation}
R_{1,\rm sine}(x)=K_{\rm sine}(x,x)=1,
\end{equation}
that is the mean level spacing is already normalised to unity. For the Bessel density we obtain after using l'H{\^ o}pital's rule
\begin{equation}
\label{Bessel-density}
R_{1,\nu}(x)=K_{\nu}(x,x)=\frac{x}{2}\left(J_\nu^2(x)-J_{\nu-1}(x)J_{\nu+1}(x)\right)\stackrel{x\to\infty}{\longrightarrow}\frac{1}{\pi},
\end{equation}
implying that the mean level spacing is $\pi$ in this normalisation of the Bessel-kernel. We will come back to this point in Section~\ref{sec:largeN}. 
The soft edge is an exception, as here the macroscopic density vanishes at the point we zoom in. However, also here is it possible to unfold the spectrum and we refer to \cite{ABKII}.

The Bessel-kernel can yield the Airy- as well as the sine-kernel in certain limits, which is rather expected. For instance, a large index $\nu$ pushes the spectral edge away from the origin so that it becomes a soft edge, see e.g. \cite{CK}, where this transition is studied on a microscopic level. This limit is not in our main focus in the present work. On the other hand, when taking the limit of large arguments in the Bessel-kernel, moving away from the hard edge, the sine-kernel is obtained. 
We are interested how much the hard edge statistics differs from the bulk statistics and will quantify this below in such a large-argument expansion.

To understand how the Bessel and the sine-kernel are related, one needs to exploit the following asymptotic expansion of the Bessel function~\cite[Eq. (10.17.3)]{NIST}
\begin{eqnarray}\label{Bessel-asymp-sine}
J_\nu(x)&=&\sqrt{\frac{2}{\pi x}}\biggl[\cos\left(x-\frac{\pi}{4}(2\nu+1)\right)
+\mathcal{O}\left(\frac{1}{x}\right)
\biggl].
\end{eqnarray}
The aim is to asymptotically expand the Bessel-kernel for $x_a,x_b\to\infty$, under the condition of a fixed difference $x_a-x_b$.
In the kernel~\eref{Bessel-kernel}, we only encounter the product of two Bessel functions, such that we consider the asymptotic expansion
\begin{eqnarray}
J_\nu( x_a)J_{\nu-1}( x_b)&=&\frac{1}{\pi\sqrt{x_ax_b}}\biggl[(-1)^\nu\cos(x_a+x_b)+\sin(x_a-x_b)
+\mathcal{O}\left(\frac{1}{x}\right) \biggl].
\end{eqnarray}
Plugging these terms into the kernel, we either have to symmetrise or anti-symmetrise these terms in $x_a$ and $x_b$. Thence, we arrive at the expansion
\begin{equation}\label{expand-intermediate}
K_\nu(x_a,x_b)=\frac{\sin(x_a-x_b)}{\pi(x_a-x_b)}-(-1)^\nu\frac{\cos(x_a+x_b)}{\pi(x_a+x_b)}+\mathcal{O}\left(\frac{1}{x_{a,b}^{2}}\right).
\end{equation}
Let us point out that we have taken $x_a,x_b\gg1$ to be of the same order. Below we will also assume that $x_a+x_b\gg|x_a-x_b|$.

The sine-kernel can be easily identified as the first term in~\eref{expand-intermediate}. The question is whether the $1/(x_a+x_b)$ can be cancelled by an unfolding transformation. For that we underline that, after unfolding, the bulk level density is  simply equal to unity, highlighting the translation invariance, cf. \cite{GMW}. In the current situation, the level density corresponding to the kernel~\eref{expand-intermediate} is equal to
\begin{eqnarray}
\rho_\nu(x_0)=\frac{1}{\pi}\biggl(1-(-1)^\nu\frac{\cos(2x_0)}{2x_0}\biggl)+\mathcal{O}\left(\frac{1}{x_0^{2}}\right).
\end{eqnarray}
We note that we understand $x_a=x_0+\delta x_a$ and $x_b=x_0+\delta x_b$, with $\delta x_a$ and $\delta x_b$ being kept fixed in the asymptotic $x_0\gg1$. The unfolding works along the formula
\begin{eqnarray}
y_a=\int_{x_0}^{x_0+\delta x_a}\rho_\nu(x')dx'=\frac{1}{\pi}\biggl[\delta x_a-(-1)^\nu\frac{\sin(2x_0+2\delta x_a)-\sin(2x_0)}{4x_0}\biggl]+\mathcal{O}\left(\delta x_a^2,\frac{1}{x_0^{2}}\right),
\end{eqnarray}
and ditto for $y_b$ and $x_b=x_0+\delta x_b$. This relation is born out of the requirement $dy_a=\rho(x_0+\delta x_a)d\delta x_a$ so that the level density in $y$ is $\rho(y)=1$, as it is for the sine-kernel.
This relation can be inverted by recursively inserting $\delta x_a$ in the non-linear part, which leads to
\begin{eqnarray}
\delta x_a=\pi y_a+(-1)^\nu\frac{\sin(2x_0+2\pi y_a)-\sin(2x_0)}{4x_0}+\mathcal{O}\left(\delta x_a^2,\frac{1}{x_0^{2}}\right).
\end{eqnarray}
From the first term we see that in this way we rescale the mean spacing of the Bessel-kernel from $1/\pi$ to unity. 
Plugging this into~\eref{expand-intermediate} and multiplying the kernel with the Jacobian of this transformation, we eventually arrive at
\begin{eqnarray}
\fl&&\frac{1}{\sqrt{\rho(x_0+\delta x_a)\rho(x_0+\delta x_b)}}K_\nu\left(x_0+\delta x_a,x_0+\delta x_b\right)\\
\fl&=&K_{\rm sine}(y_a,y_b)+(-1)^\nu\biggl[\frac{\sin[\pi(y_a-y_b)]}{\pi(y_a-y_b)}-\frac{\sin^2[\pi(y_a-y_b)]}{\pi^2(y_a-y_b)^2}-1\biggl]\frac{\cos[2x_0+\pi(y_a+y_b)]}{2x_0}+\mathcal{O}\left(\frac{1}{x_0^{2}}\right).\nonumber
\end{eqnarray}

The calculation above shows that indeed the rate of convergence of the Bessel-kernel to the sine-kernel goes with $1/x_0$ which is not very fast. Thus, it is rather amazing how close the level spacing distribution at the hard edge is to the one in the bulk, as we will see in Section~\ref{sec:comp}. It looks more like  a $1/x_0^2$ convergence than $1/x_0$. The reason is that one fixes the distance $s=y_a-y_b$ and integrates over the the center of mass $\bar{y}=y_a+y_b$. This strongly suppresses the pre-factor of the leading correction -- 
it might be equal to zero which we have not checked in the present work. Indeed, for the $2$-point correlation function, we obtain
\begin{eqnarray}
&&\frac{1}{\rho(x_0+\delta x_a)\rho(x_0+\delta x_b)}R_2\left(x_0+\delta x_a,x_0+\delta x_b\right)\\
&&=1-\frac{\sin^2[\pi(y_a-y_b)]}{\pi^2(y_a-y_b)^2}
-2(-1)^\nu\frac{\sin[\pi(y_a-y_b)]}{\pi(y_a-y_b)}\biggl[\frac{\sin[2\pi(y_a-y_b)]}{\pi(y_a-y_b)}-\frac{\sin^2[\pi(y_a-y_b)]}{\pi^2(y_a-y_b)^2}-1\biggl]
\nonumber\\
&&\quad\times\frac{\cos[2x_0+\pi(y_a+y_b)]}{2x_0}
+\mathcal{O}\left(\frac{1}{x_0^{2}}\right).\nonumber
\end{eqnarray}
When integrating over $\bar{y}=y_a+y_b$ as it has to be done for the level spacing distribution, we see that the leading order correction vanishes,
\begin{equation}
\fl\lim_{L\to\infty}\frac{1}{2L}\int_{-L}^L d\bar{y} \frac{1}{\rho(x_0+\delta x_b)\rho(x_0+\delta x_b)}R_2\left(x_0+\delta x_a,x_0+\delta x_b\right)
=1-\frac{\sin^2[\pi(y_a-y_b)]}{\pi^2(y_a-y_b)^2}+\mathcal{O}\left(\frac{1}{x_0^{2}}\right).
\end{equation}
This heuristic argument works in general for all $k$-point correlation functions, so that their average starts from $1/x_0^2$, and as the level spacing distribution between two consecutive eigenvalues is described by such an average, see Section~\ref{sec:gap}. We expect that this may explain our numerical observation of a faster convergence of the level spacing distribution from the hard edge to the bulk limit.

Since the analytical  expression which we derive in the ensuing sections are too complicated to actually see whether this argument regarding the rate of convergence is indeed the case, it remains an open problem.

Finally, let us briefly describe the limit from the Airy- to the sine-kernel. In the scaling limit leading to the Airy-kernel~\eref{Airy-kernel},  the bulk statistics is attained by taking the limit of large negative arguments. Therefore, we consider arguments $x_a=-x$, $x_b=-y$ in \eref{Airy-kernel} and take the limit $x,y\to\infty$.
The asymptotic expansion of the Airy function and its derivative read 
\begin{eqnarray}
\label{Aiasymp}
{\rm Ai}(-x)&=& \frac{1}{\sqrt{\pi}x^{1/4}}\left(\cos(\zeta-\pi/4)+ \mathcal{O}\left(\frac{1}{\zeta}\right)\right),\quad\zeta=\frac{2}{3}x^{3/2},\\
\label{Aiprimeasymp}
{\rm Ai}^\prime(-y)&=& \frac{y^{1/4}}{\sqrt{\pi}}\left(\sin(\eta-\pi/4)+ \mathcal{O}\left(\frac{1}{\eta}\right)\right), \quad\eta=\frac{2}{3}y^{3/2},
\end{eqnarray}
cf. \cite[Eqs. (9.7.9), (9.7.10)]{NIST}, respectively. Rather than starting with the unfolded Airy-kernel \cite[Eq. (VI.16)]{ABKII}, we will depart directly from \eref{Airy-kernel} and attain the sine-kernel through a change of variables. Multiplying out, we have for $x,y\to\infty$
\begin{equation}
\fl K_{\rm Airy}(-x,-y)=\frac{1}{2\pi(x-y)}\biggr[\left(\left(\frac{y}{x}\right)^{\frac14}+\left(\frac{x}{y}\right)^{\frac14}\right)\sin(\zeta-\eta)+\left(\left(\frac{y}{x}\right)^{\frac14}-\left(\frac{x}{y}\right)^{\frac14}\right)\cos(\zeta+\eta)+\mathcal{O}\left(\frac{1}{\zeta},\frac{1}{\eta}\right)\biggl].
\end{equation}
When changing variables from $x,y$ to $\zeta,\eta$, we need to take into account the Jacobian (symmetrised in both entries)
$(2/3)^{\frac13}(\zeta\eta)^{-\frac16}$, which needs to be multiplied with the kernel,
\begin{equation}
K_{\rm Airy}(\zeta,\eta)=\frac{1}{3\pi}\left[\frac{\sin(\zeta-\eta)}{(\zeta\eta)^{\frac13}(\zeta^{\frac13}-\eta^{\frac13})}-\frac{\cos(\zeta+\eta)}{(\zeta\eta)^{\frac13}(\zeta^{\frac13}+\eta^{\frac13})}\right] +\mathcal{O}\left(\frac{1}{\zeta_0^2}\right).
\label{KAirytosine}
\end{equation}
Here, the correction term is to be understood using $\zeta,\eta\approx\zeta_0\gg1$. The first term does note quite yet look like the sine-kernel. However, an expansion about the base point $\zeta_0$ yields the correct terms,
\begin{equation}
\frac{1}{3(\zeta\eta)^{\frac13}(\zeta^{\frac13}-\eta^{\frac13})}=\frac{\zeta^{\frac23}+\zeta^{\frac13}\eta^{\frac13}+\eta^{\frac23}}{3(\zeta\eta)^{\frac13}(\zeta-\eta)}\approx\frac{1}{\zeta-\eta},
\end{equation}
where only the difference between $\zeta$ and $\eta$ (or powers thereof) is of order unity. Thus, the first terms of \eref{expand-intermediate} and \eref{KAirytosine} do agree. The second term, whose prefactor is approximately
\begin{equation}
\frac{1}{3(\zeta\eta)^{\frac13}(\zeta^{\frac13}+\eta^{\frac13})}\approx\frac{1}{3\zeta_0^{\frac23}2\zeta_0^{\frac13}}=\frac{1}{6\zeta_0},
\end{equation}
can be absorbed via unfolding using the same route as starting from \eref{expand-intermediate}. Thus, we expect the same rate of convergence from the soft edge to the bulk statistics like at the hard edge.

\begin{figure}[t!]
\centerline{\includegraphics[width=0.7\textwidth]{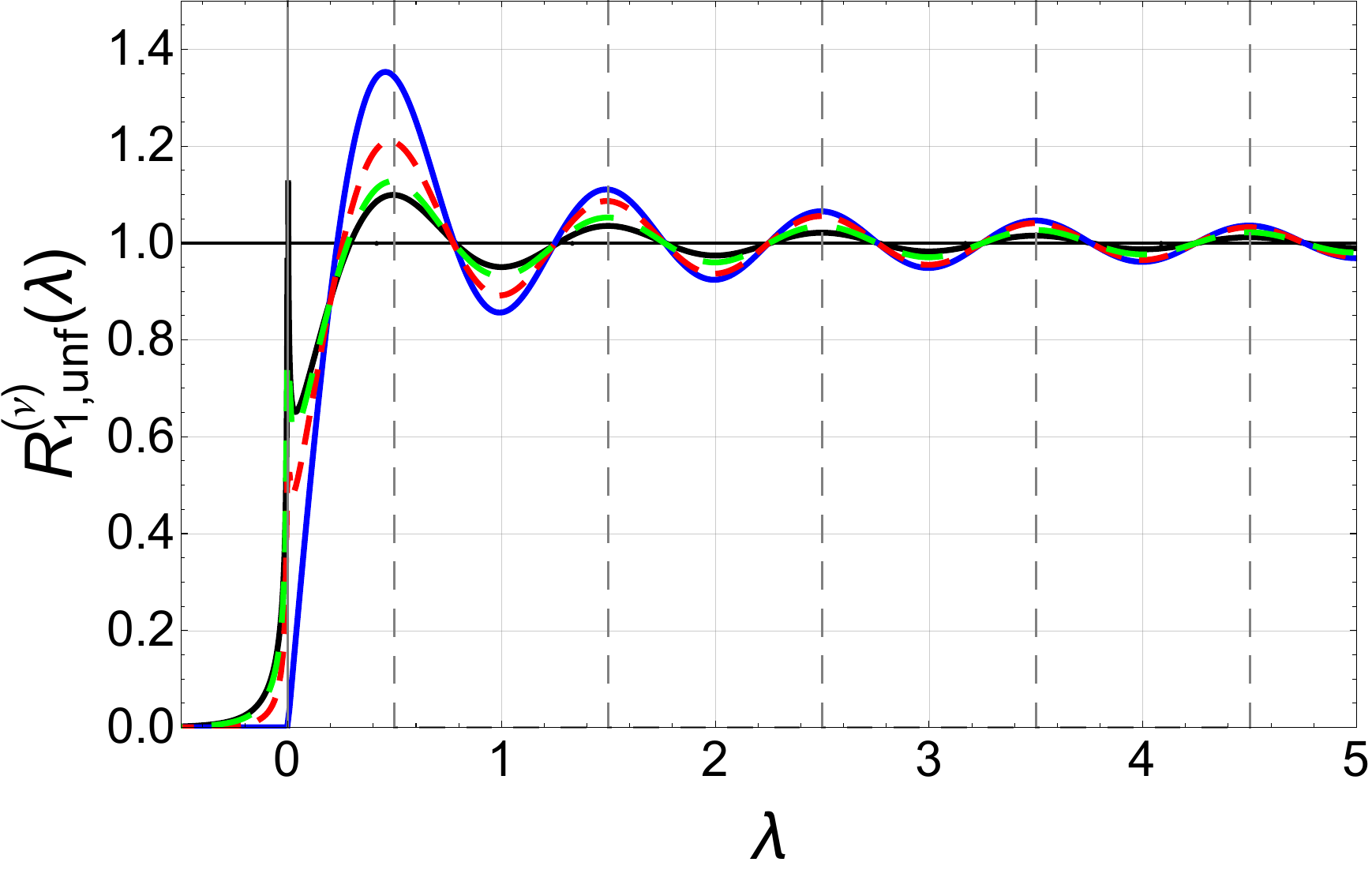}}
\caption{The properly unfolded microscopic level densities of the Bessel-kernel result~\eref{Bessel-unf} for $\nu=0$ (blue solid curve), $\nu=1$ (red finely dashed curve), $\nu=20$ (green coarsely dashed curve), and $\nu=\infty$ (black solid curve). The latter agrees with the unfolded Airy-kernel result~\eref{Airy-unf} reflected at the origin. As a guideline we have added the asymptotic unfolded level density, which is equal to the constant $1$ (solid black horizontal line) and the half-sided picket fence spectrum (spectrum of the harmonic oscillator), with eigenvalues at $n+1/2$ (dashed horizontal lines). These auxiliary lines indicate the proper unfolding of the spectra. The singularity at the origin is a relic of the unfolding, as it is not smooth at the macroscopic (as well as mesoscopic) spectral edge, cf.~\cite{ABKII}.}\label{fig:Bessel-Airy-trans}
\end{figure}

To highlight and get a feeling how close the two edge statistics are, we need to unfold both. This is particularly important as it is known~\cite{CK}, that the Bessel-kernel becomes the Airy-kernel when taking the limit $\nu\to\infty$, while the edge will be located at $x=\nu$, see~\cite{KTV}. To make both kernels, especially the Bessel-kernel with its varying parameter $\nu$ comparable, we need to properly unfold the microscopic spectrum. For the Airy-kernel~\eref{Airy-kernel}, we have already seen that the unfolding is given by $x=-\sign(\lambda)(3\pi |\lambda|/2)^{2/3}$ with $\lambda\in\mathbb{R}$. The factor $\pi$ guarantees that the mean level spacing is equal to $1$, and the minus sign reflects the spectrum at the origin so that the bulk will be to the right hand side as it is for the Bessel-kernel. The sign $\sign(\lambda)$ of $\lambda$ has to be dealt with separately, because of the root that has to be taken. Taking the limit $x_a,x_b\to -\sign(\lambda)(3\pi |\lambda|/2)^{2/3}$ by applying l'Hopital's rule, and multiplying the kernel with the Jacobian $[2\pi^2/(3 |\lambda|)]^{1/3}$, the unfolded microscopic Airy level density is given by~\cite{ABKII}
\begin{equation}\label{Airy-unf}
\fl R_{1,{\rm unf}}^{(\infty)}(\lambda)=\sign(\lambda)\left(\frac{3\pi^4|\lambda|}{2}\right)^{1/3}{\rm Ai}^2\left(-\sign(\lambda)\biggl[\frac{3\pi|\lambda|}{2}\biggl]^{2/3}\right)+\left(\frac{2\pi^2}{3|\lambda|}\right)^{1/3}{{\rm Ai}'}^2\left(-\sign(\lambda)\biggl[\frac{3\pi|\lambda|}{2}\biggl]^{2/3}\right).
\end{equation}
The superscript $\infty$ indicates that this expression agrees with the properly unfolded Bessel-kernel density~\eref{Bessel-density} in the limit $\nu\to\infty$. Evidently there will be a singularity at the origin, which is a relict of the unfolding and has no deeper meaning, cf., Figure~\ref{fig:Bessel-Airy-trans}.

To get the proper unfolding for all $\nu$ of the hard edge statistics, we start from the integral representation of the Bessel function
\begin{equation}
J_\nu(x)=\int_{-\pi}^\pi\exp[ix\sin(\varphi)-i\nu\varphi]\frac{d\varphi}{2\pi},
\end{equation}
for an arbitrary integer $\nu\in\mathbb{N}_0$. For large $x\gg1$ and $\nu\gg1$, regardless what kind of relation they have, it is clear that the saddle point equation gives $x\cos(\varphi)=\nu$. This can be only solved when $x\geq\nu$, hinting at the creation of a soft edge. In the regime $x\geq\nu$, via standard stationary phase approximation the Bessel function becomes as follows
\begin{equation}
J_\nu(x)\approx\frac{1}{\pi}\sqrt{\frac{2}{x}}\frac{1}{(1-\nu^2/x^2)^{1/4}}\cos\left[x\sqrt{1-\frac{\nu^2}{x^2}}-\nu\,{\rm arccos}\left(\frac{\nu}{x}\right)-\frac{\pi}{4}\right].
\end{equation}
 In particular limits, this agrees with the asymptotic in~\cite[Theorem~1.3]{CK}. A simple comparison with~\eref{Bessel-asymp-sine} shows that the proper unfolding is equal to
\begin{equation}
\lambda=g(x)= \frac{1}{\pi}\left[x\sqrt{1-\frac{\nu^2}{x^2}}-\nu\,{\rm arccos}\left(\frac{\nu}{x}\right)\right],\quad{\rm for}\ x\geq\nu.
\end{equation} 
As a check, the Taylor expansion gives the behaviour $\lambda\propto (x/\nu-1)^{3/2}$ in the limit $\nu/x\approx 1$, which is the unfolding of the Airy-kernel. To mimic the sign $\sign(\lambda)$ that has appeared for the unfolding of the soft edge statistics, we choose
\begin{equation}
\lambda=g(x)=\frac{1}{\pi}\left[x\sqrt{\frac{\nu^2}{x^2}-1}-\nu\,{\rm arccosh}\left(\frac{\nu}{x}\right)\right],\quad{\rm when}\ x<\nu.
\end{equation}
One can check that $g(x)$ is monotonously increasing on $\mathbb{R}_+$. Indeed again a Taylor expansion about $\nu/x\approx 1$ gives the proper unfolding of the soft edge statistics. To invert the relation between $x$ and $\lambda$ we have exploited the integral formula
\begin{equation}
x(\lambda)=\int_0^\infty\Theta[\lambda-g(\tilde{x})]d\tilde{x},
\end{equation}
where $\Theta$ is the Heaviside step function.
The Jacobian of this substitution is
\begin{equation}
\frac{dx}{d\lambda}(\lambda)=\left\{\begin{array}{cl} \displaystyle \frac{\pi}{\sqrt{1-\nu^2/x^2(\lambda)}}, & x(\lambda)\geq\nu, \\ \displaystyle \frac{\pi}{\sqrt{\nu^2/x^2(\lambda)-1}},  & x(\lambda)<\nu.\end{array}\right.
\end{equation}
Thus, the unfolded Bessel-kernel is given by 
\begin{equation}\label{Bessel-unf}
R_{1,{\rm unf}}^{(\nu)}(\lambda)=\frac{x(\lambda)}{2}\left[J_\nu^2(x(\lambda))-J_{\nu-1}(x(\lambda))J_{\nu+1}(x(\lambda))\right]\frac{dx}{d\lambda}(\lambda).
\end{equation}
It indeed satisfies the point-wise limit $\lim_{\nu\to\infty}R_{1,{\rm unf}}^{(\nu)}(\lambda)=R_{1,{\rm unf}}^{(\infty)}(\lambda)$ for fixed $\lambda$, see~\eref{Airy-unf} and Figure~\ref{fig:Bessel-Airy-trans}.

The unfolded microscopic  level densities~\eref{Airy-unf} and~\eref{Bessel-unf} are illustrated in Figure~\ref{fig:Bessel-Airy-trans}. All curves from $\nu=0$ to $\nu=\infty$ almost perfectly agree, starting from the fourth eigenvalue. Hence we would not expect a big difference in any of the level spacing distributions. For $\nu=20$ the Bessel-kernel seems to agree with the Airy-kernel even from the first eigenvalue, judged with the bare eye. Hence, we also expect a quick convergence in increasing $\nu$. The biggest deviation for the level spacing distribution might be seen between the first and second eigenvalue for $\nu=0$.

\section{Gap probabilities and Level Spacing Distributions}\label{sec:gap}

Let us underscore that most of the ensuing discussion in the present section holds for arbitrary positive real spectra with a finite number of eigenvalues. Thus our particular choice of an ensemble can be readily generalised.

We start from a generic joint probability density $P_n(\{x\})$ of the set of $n$ unordered eigenvalues 
 $\{x\}=\{x_i\}_{i=1}^n$, with $x_i\in\mathbb{R}_+$ for all $i=1,\ldots,n$. The emphasis on ``unordered'' is important, as some combinatorial constants depend on whether the eigenvalues have been ordered or not. In our case, the normalised joint probability density of the  $\chi$GUE~\cite{SV93,Jac3fold}  with $N_{\rm f}$ inserted characteristic polynomials is given by
\begin{equation}\label{QCD-jpdf}
P_{\nu,n}^{(\Nf)}(\{x\})=\frac{1}{Z_{\nu,n}^{(N_{\rm f})}(\{m\})}\Delta_n\left(\{x^2\}\right)^2\prod_{j=1}^n \left(x_j^{2\nu+1}e^{-x_j^2}\prod_{f=1}^{N_{\rm f}}(x_j^2+m_f^2)\right), 
\end{equation}
with the Vandermonde determinant defined as $\Delta_n(\{x\})=\prod_{1\leq a<b\leq n}(x_b-x_a)$. This density depends on several external parameters: $N_{\rm f}$ masses $\{m\}=\{m_f\}_{f=1}^{N_{\rm f}}$,  the quark masses in applications to QCD, and the topological charge $\nu\in\mathbb{N}$ counting the number of zero eigenvalues of the Dirac operator (rectangularity of the $n\times(n+\nu)$ random matrix). 
For more details in this relation to QCD we refer to \cite{reviewQCD}.
The normalisation constant (partition function) is denoted by 
\begin{equation}
Z_{\nu,n}^{\Nf}(\{m\})=\int_{[0,\infty)^n}dx_1\ldots dx_{n}\Delta_n\left(\{x^2\}\right)^2\prod_{j=1}^n \left(x_j^{2\nu+1}e^{-x_j^2}\prod_{f=1}^{N_{\rm f}}(x_j^2+m_f^2)\right).
\end{equation}

As already mentioned, one common technique to compute the level spacing distribution is via gap probabilities using Fredholm determinants, see~\cite{Handbook}. These yield the probability that a certain interval is void of eigenvalues.
As we aim for  
the level spacing distribution between a specific pair of consecutive eigenvalues,  we define the $k$th gap probability as follows. It implements the condition that $k$ eigenvalues lie below the gap, and the remaining  $n-k$ eigenvalues are above the gap. When the gap is  the interval $[a,b]\in\mathbb{R}_+$, this conditional gap probability is defined by the $n$-fold integral
\begin{equation}\label{gap-prob}
 E_k([a,b])=\frac{n!}{k!(n-k)!}\int_{[0,a]^{k}}dx_1\ldots dx_{k}\int_{[b,\infty)^{n-k}}dx_{k+1}\ldots dx_nP_n(\{x\}).
\end{equation}
The combinatorial pre-factor takes care of the fact that the eigenvalues are unordered and ensures the proper normalisation. 
For $k=0$, only the second set of integrals is present, that is all eigenvalues are larger than or equal to $ b$. The gap probability can be also expressed in terms of the kernel of the corresponding orthogonal polynomials, that is the generalised Laguerre polynomials for the $\chi$GUE at $\Nf=0$, see Section \ref{sec:finiteN}. However, we will directly compute the spacing distribution in a different way to be introduced now, circumventing the problem to determine the orthogonal polynomials when introducing $\Nf$ flavours in the weight function.

When taking the mixed second derivative of  Eq.~\eref{gap-prob} in $a$ and $b$ for $k\geq1$, we obtain the joint density that the $k$th eigenvalue sits at the position $a$ and the $(k+1)$st eigenvalue at the position $b$. Defining the spacing $s=b-a$, and integrating over their starting point $a=x_{k}$, yields the level spacing distribution between the $k$th and $(k+1)$st eigenvalue,
\begin{eqnarray}\label{level-space.a}
 p_{k,n}(s)&=&\frac{n!}{(k-1)!(n-k-1)!}\int_{0}^\infty dx_{k}\int_{[0,x_{k}]^{k-1}}dx_1\ldots dx_{k-1}\\
&&\times \int_{[x_{k}+s,\infty)^{n-k-1}}dx_{k+2}\ldots dx_nP_n(x_1,\ldots,x_{k},x_{k}+s,x_{k+2},\ldots,x_n).\nonumber
\end{eqnarray}
Here, we explicitly spell out the arguments of the joint density $P_n(\{x\})$. When $k=1$ the set of integrals over $[0,x_{k}]$ is absent. It can be shown that the spacing distribution is properly normalised for all $k$, $\int_0^\infty ds  p_{k,n}(s)=1$, cf.~\cite[Appendix A]{AD03}. 

Our aim is to study the spacing for the smallest eigenvalues. Therefore, we fix the smallest $k+1$ in the large $n$-limit. This means that the integral over the remaining $n-k-1$ eigenvalues can be understood as being proportional to 
a partition function of $n-k-1$ eigenvalues, depending on an extended number of shifted masses. 
Indeed, a standard trick~\cite{DN} is to define new variables $y_j=\sqrt{x_{k+1+j}^2-(x_{k}+s)^2}$ for $j=1,\ldots,n-k-1$. These can be interpreted as eigenvalues of a positive definite Hermitian random matrix of dimension $(n-k-1)\times(n-k-1+2)$. The expression for the level spacing distribution in terms of the shifted variables only changes marginally
\begin{eqnarray}
 \fl p_{k,n}(s)&=&\frac{n!}{(k-1)!(n-k-1)!}\int_{0}^\infty dx_{k}\int_{[0,x_{k}]^{k-1}}dx_1\ldots dx_{k-1} \int_{[0,\infty)^{n-k-1}}
 \prod_{j=1}^{n-k-1}\frac{dy_jy_j}{\sqrt{y_j^2+(x_{k}+s)^2}}
 \nonumber\\
\fl&&\times P_{n}\left(x_1,\ldots,x_{k},x_{k}+s,\sqrt{y_1^2+(x_{k}+s)^2},\ldots,\sqrt{y_{n-k-1}^2+(x_{k}+s)^2}\right).\label{level-space.b}
\end{eqnarray}
Once again we have explicitly spelled out the arguments of the joint probability density.

At first glance, the expression~\eref{level-space.b} looks more cumbersome than the one we started from. Yet, for the joint probability density \eref{QCD-jpdf} of the $\chi$GUE with $\Nf$ flavours this is a considerable simplification, because it relates two joint probability densities of the same random matrix ensemble, but for different matrix sizes and different numbers of masses. In particular, using that the Vandermonde determinant is invariant under translation of its variables, and the fact that it can be split  in terms of two sets of variables as follows,
\begin{equation}
\fl\Delta_n(x_1,\ldots,x_{k+1},z_1,\ldots,z_{n-k-1})=\Delta_{k+1}(x_1,\ldots,x_{k+1})\Delta_{n-k-1}(z_1,\ldots,z_{n-k-1})
\prod_{i=1}^{k+1}\prod_{j=1}^{n-k-1}(z_j-x_i),
\end{equation}
it holds that
\begin{eqnarray}
&&P_{\nu,n}^{(\Nf)}\left(x_1,\ldots,x_{k},x_{k}+s,
\sqrt{y_1^2+(x_{k}+s)^2},\ldots,\sqrt{y_{n-k-1}^2+(x_{k}+s)^2} \right)
\prod_{j=1}^{n-k-1}\frac{y_j}{\sqrt{y_j^2+(x_{k}+s)^2}}
\nonumber\\
&&=\frac{Z_{2,n-k-1}^{(\tNf)}(\{\widetilde{m}\})}{Z_{\nu,n}^{(\Nf)}(\{m\})}\Delta_{k+1}\left(x_1^2,\ldots,x_{k}^2,(x_{k}+s)^2\right)^2(x_{k}+s)^{2\nu+1}e^{-(n-k)(x_{k}+s)^2}\nonumber\\
&&\quad\times\left(\prod_{f=1}^{N_{\rm f}}((x_{k}+s)^2+m_f^2)\right)\prod_{j=1}^{k} \left(x_j^{2\nu+1}e^{-x_j^2}\prod_{f=1}^{N_{\rm f}}(x_j^2+m_f^2)\right)P_{2,n-k-1}^{(\tNf)}\left(\{y\}\right).\label{new-jpdf}
\end{eqnarray}
Here, we have defined an extended number of flavours 
\begin{equation}
\tNf=\Nf+\nu+2k.
\end{equation}
The enlarged set of $\tNf$ (shifted) mass parameters is reading
\begin{eqnarray}\label{mass-new}
\{\widetilde{m}\}&=&\left\{x_{k}+s,\ldots,x_{k}+s,
\sqrt{m_1^2+(x_{k}+s)^2},\ldots,\sqrt{m_{\Nf}^2+(x_{k}+s)^2},\right.\\
&&\left.
\sqrt{(x_{k}+s)^2-x_1^2},\sqrt{(x_{k}+s)^2-x_1^2},\ldots,
\sqrt{(x_{k}+s)^2-x_{k}^2},\sqrt{(x_{k}+s)^2-x_{k}^2}
\right\}.\nonumber
\end{eqnarray}
The set of new masses $(x_{k}+s)$ is $\nu$-fold degenerate, whereas the masses $\sqrt{(x_{k}+s)^2-x_j^2}$ are two-fold degenerate for $j=1,\ldots,k$. Obviously the last two masses can be simplified to $\sqrt{s(2x_{k}+s)}$.
Notice that the shift has promoted the $\nu$ zero-eigenvalues in \eref{QCD-jpdf} to become degenerate mass-terms, whereas the squared Vandermonde determinant has created a fixed number of zero-eigenvalues $\nu=2$ in the new variables $y_1,\ldots,y_{n-k-1}$.

This relation has been exploited several times in the literature~\cite{GWW,DNW,DN}. In our case, it creates a short-cut as the pre-factor in front of the joint probability density $P_{2,n-k-1}^{(\tNf)}\left(\{y\}\right)$, the ratio of two partition functions, is $y$-independent. Hence, the integral of $P_{2,n-k-1}^{(\tNf)}\left(\{y\}\right)$ over all its eigenvalues $y_1,\ldots,y_{n-k-1}$ yields unity, so that we end up with a $k$-fold integral for the level spacing distribution of the $\chi$GUE with $\Nf$ flavours
\begin{eqnarray}
 \fl p_{k,n}^{(\nu,\Nf)}(s)&=&\frac{n!}{(k-1)!(n-k-1)!}\int_{0}^\infty dx_{k}\int_{[0,x_{k}]^{k-1}}dx_1\ldots dx_{k-1} \frac{Z_{2,n-k-1}^{(\tNf)}(\{\widetilde{m}\})}{Z_{\nu,n}^{(\Nf)}(\{m\})}\Delta_{k}\left(x_1^2,\ldots,x_{k}^2\right)^2(x_{k}+s)^{2\nu+1}\nonumber\\
\fl&&\times e^{-(n-k)(x_{k}+s)^2}\prod_{f=1}^{N_{\rm f}}((x_{k}+s)^2+m_f^2)\prod_{j=1}^{k} \left([(x_{k}+s)^2-x_j^2]^2x_j^{2\nu+1}e^{-x_j^2}\prod_{f=1}^{N_{\rm f}}(x_j^2+m_f^2)\right).\label{level-space.c}
\end{eqnarray}
Notice that the last line is proportional to $s^2$, due to the term with $j=k$.
Certainly, part of the difficulty has been moved into the evaluation of the new partition function $Z_{2,n-k-1}^{(\tNf)}(\{\widetilde{m}\})$. Yet, its explicit form is known in principle  and will be spelled out  in the following section.

One last remark is in order which applies to arbitrary ensembles. The level spacing distribution~\eref{level-space.b}  still has to be unfolded, meaning its first moment
\begin{equation}\label{first-moment}
\bar{s}_{k,n}=\int_0^\infty ds\,s\, p_{k,n}(s), 
\end{equation}
does not necessarily equal to $1$ yet. This is especially the case for the result~\eref{level-space.c}. Under the assumption that the macroscopic or mesoscopic level density is already in a constant form (the microscopic level density asymptotes to a constant when taking the limit into the bulk), we only need to rescale the spacing distribution to make it comparable with standard distributions such as the Wigner surmise~\eref{Wignersurmise}, where the first moment is unity. This means, one needs to consider the rescaled spacing distribution
\begin{equation}\label{unfolded-level-density}
\bar{p}_{k,n}(s)=
\bar{s}_{k,n} p_{k,n}\left(\bar{s}_{k,n} s\right).
\end{equation}
Note that when we rescale the spacing between two eigenvalues we also have to rescale all other eigenvalues, as well as the masses $m_j$ with the same scaling factor.
For practical purposes one can also compute the mean spacing $\bar{s}_{k,n}$ through the difference of the mean positions of the $k$th and $(k+1)$st smallest eigenvalue.


\section{Analysis at Finite Matrix Dimension}\label{sec:finiteN}

\subsection{Orthogonal polynomials and partition function}

To compute the partition functions $Z_{\nu,n}^{(\Nf)}(\{m\})$ and $Z_{2,n-k-1}^{(\tNf)}(\{\widetilde{m}\})$ of the $\chi$GUE with a given number of flavours,
we begin with the $\chi$GUE at $\Nf=0$, where the partition function and all $k$-point correlation functions can be expressed in terms of 
the generalised Laguerre polynomials and their kernel. Their orthogonality relation
with squared norms $h_j$ reads \cite[Table 18.3.1]{NIST} 
\begin{equation}
\int_0^\infty dz L_{l}^{(\nu)}(z^2)L_{j}^{(\nu)}(z^2)z^{2\nu+1}e^{-z^2}=
h_j\delta_{lj}, \qquad h_j=\frac{(j+\nu)!}{j!\,2}\, ,
\end{equation}
valid for $\nu>-1$ for convergence.  Here~\cite[Eq.~(8.5.12)]{NIST}
\begin{equation}\label{Laguerre-def}
L_{j}^{(\nu)}(z)=\sum_{l=0}^{j}\frac{(j+\nu)!}{j!(j-l)!\Gamma(\nu+l+1)}(-z)^l
\end{equation}
is the definition of the generalised Laguerre polynomials in their standard normalisation. Notice that in this definition the parameter $\nu$ can be continued to negative integer values, meaning that the sum will be cut off from below, to start at $l>-\nu$ only. For later convenience we also introduce generalised Laguerre polynomials in monic normalisation, $\widehat{L}_n^{(\nu)}(z)=z^n+O(z^{n-1})$, with its corresponding norms
\begin{equation}
\widehat{L}_n^{(\nu}(z)=(-1)^nn!\,L_n^{(\nu)}(z),\quad \hat{h}_j=j!(j+\nu)!/2.
\label{L-monic}
\end{equation}
The corresponding kernel determines all $k$-point correlation functions of the $\chi$GUE with $\Nf=0$ through \eref{k-point-det}  at finite-$n$, 
\begin{equation}
\label{k-point-det.n}
R_{k,n}(x_1,\ldots,x_k)=\prod_{c=1}^k x_c^{2\nu+1}e^{-x_c^2}
\det[K_{\nu,n}^{(\chi {\rm GUE})}(x_a^2,x_b^2)]_{a,b=1,\ldots,k}.
\end{equation}
In contrast to Section \ref{sec:heuristic}, we use here squared variables for the (pre-)kernel $K_{\nu,n}^{(\chi {\rm GUE})}(x_a^2,x_b^2)$, as we later have to differentiate with respect to these. Also note that we have not included the weight function into the kernel. 
It only contains the orthogonal polynomials and is given by 
\begin{eqnarray}\label{Laguerre-kernel}
\fl  K_{\nu,n}^{(\chi {\rm GUE})}(x_1^2,x_2^2)&=&
 \sum_{j=0}^{n-1} 2\frac{j!}{(j+\nu)!}L_{j}^{(\nu)}(x_1^2)L_{j}^{(\nu)}(x_2^2)=2\frac{n!}{(n+\nu-1)!}
\frac{L_{n-1}^{(\nu)}(x_1^2)L_{n}^{(\nu)}(x_2^2)-L_{n}^{(\nu)}(x_1^2)L_{n-1}^{(\nu)}(x_2^2)}{x_1^2-x_2^2},
\end{eqnarray}
where in the second equality we have used the Christoffel-Darboux identity for $x_1\neq x_2$, see~\cite[Eq.~18.2.12]{NIST}.
Note that the determinant~\eref{k-point-det.n} is invariant under a rescaling of the kernel $K(x_a^2,x_b^2)\to g(x_a^2)/g(x_b^2)$ $K(x_a^2,x_b^2)$ for any non-zero function $g(x)$. Two kernels related in this way are called equivalent kernels, leading to the same $k$-point correlation functions and, thence, the same spectral statistics.

When $x_1=x_2=x$ we need to apply l'H\^opital's rule to \eref{Laguerre-kernel}, cf. \cite[Eq.~18.2.13]{NIST},  
\begin{eqnarray}
 K_{\nu,n}^{(\chi {\rm GUE})}(x^2,x^2)=2\frac{n!}{(n+\nu-1)!}
 \left( L_{n-1}^{(\nu+1)}(x^2)L_{n-1}^{(\nu)}(x^2)-L_{n-2}^{(\nu+1)}(x^2)L_{n}^{(\nu)}(x^2)\right),
\label{level-den-finite}
\end{eqnarray}
upon using ~\cite[Eq.~(18.9.23)]{NIST} 
\begin{equation}\label{Lag-derivative}
{\partial_z} L_j^{(\nu)}(z)=-L_{j-1}^{(\nu+1)}(z),
\end{equation}
in terms of squared variables.
The recurrence relation~\cite[Eq.~(18.9.13)]{NIST},
\begin{equation} \label{Lag-recurrence}
L_j^{(\nu)}(z)=L_j^{(\nu+1)}(z)-L_{j-1}^{(\nu+1)}(z),
\end{equation}
is employed  to bring
~\eref{level-den-finite} into the form
\begin{equation}
 K_{\nu,n}^{(\chi {\rm GUE})}(x^2,x^2)=\frac{n!}{(n+\nu-1)!}2
 \left( L_{n}^{(\nu)}(x^2)L_{n-1}^{(\nu)}(x^2)-L_{n-1}^{(\nu+1)}(x^2)L_{n}^{(\nu-1)}(x^2)\right).
\label{level-den-finite.b}
\end{equation}
In this form, compared to \eref{level-den-finite}, the leading order asymptotic of the Laguerre polynomials for large-$n$ is no longer cancelling and thus non-vanishing.
Equation~\eref{level-den-finite.b} implies in particular the following expression for the spectral density at finite matrix dimension $n$,
$R_{1,n}(x)= x^{2\nu+1}e^{-x^2}K_{\nu,n}^{(\chi {\rm GUE})}(x^2,x^2)$ from \eref{k-point-det.n}. It is normalised to the number of points $n$.

When interpreting the $\Nf$ flavours added to the $\chi$GUE in \eref{QCD-jpdf} as expectation values of characteristic polynomial in the $\chi$GUE at $\Nf=0$, the following expression is well known, see e.g. \cite{reviewQCD} for a derivation,
\begin{eqnarray}
 Z_{\nu,n}^{(\Nf)}(\{m\})= n!\left(\prod_{j=0}^{N_{\rm f}-1} (n+j)!\right)\left(\prod_{j=0}^{n-1}\frac{j!(j+\nu)!}{2}\right)\frac{\det\left[L_{n+b-1}^{(\nu)}(-m_a^2)\right]_{a,b=1,\ldots,N_{\rm f}}}{\Delta_{N_{\rm f}}(\{m^2\})}.\label{Zdenom}
\end{eqnarray}
The factor $n!$ in front reflects that the eigenvalues have not been ordered. 
For $\Nf=0$ the first product and last ratio are absent, and we recover
\begin{equation}
Z_{\nu,n}^{(\Nf=0)}=n!\prod_{j=0}^{n-1}\hat{h}_j.
\label{ZNf0-norm}
\end{equation}
The preparation of~\eref{Zdenom} for the large $n$-analysis to be done in Section~\ref{sec:largeN} is based on the relation~\cite[Eq.~(18.9.14)]{NIST}
\begin{equation}\label{Lag-recurrence.b}
L_j^{(\nu)}(z)=-\frac{z}{j}L_{j-1}^{(\nu+1)}(z)+\frac{j+\nu}{j}L_{j-1}^{(\nu)}(z)\quad{\rm for}\ j>0,
\end{equation}
applied recursively. Taking linear combinations of columns in the determinant of~\eref{Zdenom} allows us to rewrite \eref{Zdenom} as
\begin{eqnarray}
 Z_{\nu,n}^{(\Nf)}(\{m\})= (n!)^{N_{\rm f}+1}\left(\prod_{j=0}^{n-1}\frac{j!(j+\nu)!}{2}\right)\frac{\det\left[m_a^{2b-2} L_{n}^{(\nu+b-1)}(-m_a^2)\right]_{a,b=1,\ldots,N_{\rm f}}}{\Delta_{N_{\rm f}}(\{m^2\})}.\label{Zdenom.b}
\end{eqnarray}
Now, the columns no longer become degenerate to leading order in the large-$n$ asymptotic, as they would have done in \eref{Zdenom}. 
The normalisation of \eref{Zdenom.b} can be readily checked in the quenched limit ${m_1,\ldots,m_{N_{\rm f}}\to\infty}$.

\subsection{Partition function of shifted masses}

In order to determine the second partition function 
$Z_{2,n-k-1}^{(\tNf)}(\{\widetilde{m}\})$ with an extended number of shifted masses \eref{mass-new}, which are partly degenerate, we will use the following theorem from \cite{AV}. It expresses the expectation value of the product of characteristic polynomials in several equivalent forms, in terms of the determinant of a matrix of kernels and orthogonal polynomials. Adapted to our setting of the $\chi$GUE, it reads for $K\geq L$
\begin{eqnarray}
\fl&&\frac{1}{Z_{\nu,n}^{(0)}}
\int_{[0,\infty)^n}\prod_{j=1}^n\left(dy_jy_j^{2\nu+1}e^{-y_j^2}\prod_{k=1}^K(v_k^2-y_j^2)\prod_{l=1}^L(u_l^2-y_j^2)\right)
\Delta_n(\{y^2\})^2\nonumber\\
\fl&=& \frac{\prod_{j=n}^{n+L-1}\hat{h}_j}{\Delta_K(\{v^2\})\Delta_L(\{u^2\})}
\det\left[\begin{array}{c|c}  
K_{\nu,n+L}^{(\chi {\rm GUE})}(v_a^2,u_b^2) & 
\widehat{L}_{n+L}^{(\nu)}(v_a^2),\ \ldots,\ \widehat{L}_{n+K-1}^{(\nu)}(v_a^2)\\
\end{array}\right]_{\tiny\begin{array}{l} a=1,\ldots,K\\b=1,\ldots,L\end{array}}.
\label{ThmAV}
\end{eqnarray}
Notice that in \cite{AV} the expression on the right hand side is given here in terms of orthogonal polynomials in monic normalisation and their norms \eref{L-monic}. 
The point is that we may split the $K+L$ characteristic polynomials or masses in two groups as we like, resulting in many equivalent forms of determinants of different size. In \cite{AV}, valid for Hermitian and non-Hermitian ensembles of random matrices, this split was introduced when considering the product of $K$ characteristic polynomials and $L$ complex conjugated ones  in the non-Hermitian case.

Considering the $\tNf$ new masses in \eref{mass-new}, it is unavoidable to apply l'H{\^o}pital's rule to the $\nu$-fold degenerate mass $x_{k}+s$. However, some simplification can be achieved by dividing the $k$ two-fold degenerate masses into the two groups of characteristic polynomials with parameters $\{v^2\}$ and $\{u^2\}$, thus choosing $K=\Nf+\nu+k$ and $L=k$ in~\eref{ThmAV}.
Before taking into account this degeneracy, let us take one more step of preparation, in sending $v_i^2\to-v_i^2$ and $u_j^2\to-u_j^2$ in \eref{ThmAV}, to achieve the correct signs in the mass terms, by switching polynomials and kernels inside the determinant, and by moving from monic to standard generalised Laguerre polynomials
\begin{eqnarray}
&&(-1)^{n(K+L)}\int_{[0,\infty)^n}\prod_{j=1}^n\left(dy_jy_j^{2\nu+1}e^{-y_j^2}\prod_{k=1}^K(v_k^2+y_j^2)\prod_{l=1}^L(u_l^2+y_j^2)\right)
\Delta_n(\{y^2\})^2\nonumber\\
&=&\frac{(-1)^{L(K-L)+(K-L)(n+L+(K-L-1)/2)}n!\prod_{j=0}^{n+L-1}\frac{j!(j+\nu)!}{2}\prod_{l=n+L}^{n+K-1}l!
}{
(-1)^{\frac{K(K-1)}{2}+\frac{L(L-1)}{2}}
\Delta_K(\{v^2\})\Delta_L(\{u^2\})}\nonumber\\
&&\times
\det\left[\begin{array}{c|c}  
L_{n+L}^{(\nu)}(-v_a^2),\ \ldots,\ L_{n+K-1}^{(\nu)}(-v_a^2)&K_{\nu,n+L}^{(\chi {\rm GUE})}(-v_a^2,-u_b^2)\\
\end{array}\right]_{\tiny\begin{array}{l} a=1,\ldots,K\\b=1,\ldots,L\end{array}},
\label{ThmAV-}
\end{eqnarray}
where we have multiplied by \eref{ZNf0-norm}. Regarding the degeneracy of some of the parameters $v_j$, it is not difficult to see that for any set of suitably differentiable functions $f_1,\ldots,f_\nu$,  the following limit applies:
\begin{eqnarray}
\lim_{z_1,\ldots,z_\nu\to x}\frac{1}{\Delta_\nu(\{z\})}\det[f_a(z_b)]_{a,b=1,\ldots,\nu}=\prod_{j=0}^{\nu-1}\frac{1}{j!} \det[f_a^{(b-1)}(x)]_{a,b=1,\ldots,\nu}, 
\end{eqnarray}
where $f_a^{(b)}(x)$ denotes the $b$th derivative with respect to the argument. The size of the determinant can be trivially extended, as long as the first $\nu$ rows depend in the same way on the arguments $z_b$ and thus become degenerate in the limit. 

Let us turn to the evaluation of the partition function $Z_{2,n-k-1}^{(\tNf)}(\{m\})$. In order to apply~\eref{ThmAV-}, we split the $\tNf=N_{\rm f}+\nu+2k$ masses into two sets, which is born out of~\eref{mass-new},
\begin{eqnarray}
&&v_1=\ldots=v_\nu=x_{k}+s\equiv\widetilde{s};\quad v_{\nu+j}=\sqrt{m_f^2+(x_{k}+s)^2}\equiv\widetilde{m}_f\quad \mbox{for}\ f=1,\ldots,\Nf; \nonumber\\
&&v_{\nu+\Nf+j}=u_j= \sqrt{(x_{k}+s)^2-x_j^2}\equiv\widetilde{x}_j\quad\mbox{for}\ j=1,\ldots,k.
\label{shift-var}
\end{eqnarray}
Notice that in the last term for $j=k$, we obtain
\begin{equation}
v_{\nu+\Nf+k}=u_{k}= \sqrt{s(2x_{k}+s)}=\widetilde{x}_{k},
\end{equation}
which is proportional to $\sqrt{s}$.
These are the variables that will appear inside the polynomials and kernels on the right hand side of \eref{ThmAV-}. In the Vandermonde determinants of parameters $\{u^2\}$ and $\{v^2\}$, due to their translational invariance, the following simplifications occur:
\begin{equation}
\Delta_{k}(\{u^2\})=(-1)^{\frac{k(k-1)}{2}}\Delta_{k}(\{x^2\}),
\end{equation}
and 
\begin{equation}
\Delta_{\Nf+\nu+k}(\{v^2\})=\Delta_{\nu}(v_1^2,\ldots,v_\nu^2)
\Delta_{\Nf+k}(v_{\nu+1}^2,\ldots,v_{\nu+\Nf+k}^2)\prod_{i=1}^\nu\prod_{j=1}^{\Nf+k}(v_{\nu+j}^2-v_i^2),
\end{equation}
with 
\begin{eqnarray}
\Delta_{\Nf+k}(v_{\nu+1}^2,\ldots,v_{\nu+\Nf+k}^2)&=&(-1)^{\frac{k(k-1)}{2}+\Nf k}\Delta_{\Nf}(\{m^2\})\Delta_{k}(\{x^2\})\prod_{j=1}^{k}\prod_{f=1}^{\Nf}(m_f^2+x_j^2),\nonumber\\
\prod_{i=1}^\nu\prod_{j=1}^{\Nf+k}(v_{\nu+j}^2-v_i^2)&=&(-1)^{\nu k}\prod_{f=1}^{\Nf} m_f^{2\nu}\prod_{j=1}^{k}x_j^{2\nu}.
\end{eqnarray}
In the last line, we have already applied the degeneracy of the first $\nu$ masses.
Before taking the degenerate limit of the first $\nu$ parameters $v_i$, we multiply the first $\nu$ rows of the determinant by $\widetilde{s}^4$ for later convenience, and thus divide the pre-factor by $\widetilde{s}^{4\nu}$.
Putting all this together, we obtain the following expression for the partition function of $\tNf$ shifted masses with $n-k-1$ eigenvalues and 2 zero-modes
\begin{eqnarray}
 \fl Z_{2,n-k-1}^{(\tNf)}(\{\widetilde{m}\})&=&\frac{2^{1-n}(n-k-1)!\prod_{j=0}^{n-2}j!(j+2)!\prod_{l=n-1}^{\Nf+\nu+n-2}l!}{\widetilde{s}^{4\nu}\prod_{j=0}^{\nu-1}j!\Delta_{k}(\{x^2\})^2\Delta_{\Nf}(\{m^2\})\left[\prod_{f=1}^{\Nf}m_f^{2\nu}\prod_{j=1}^{k}x_j^{2\nu}(m_f^2+x_j^2)\right]} \nonumber\\
\fl&&\times\det\left[\begin{array}{c|c} 
\partial^{a-1}_{\tilde{s}^2}\widetilde{s}^4L_{n+d-2}^{(2)}(-\widetilde{s}^2) & 
\displaystyle \partial_{\tilde{s}^2}^{a-1}\widetilde{s}^4 K_{2,n-1}^{(\chi {\rm GUE})}(-\widetilde{s}^2,-\widetilde{x}_e^2)\\
\hline L_{n+d-2}^{(2)}(-\widetilde{m}_b^2) & 
\displaystyle K_{2,n-1}^{(\chi {\rm GUE})}(-\widetilde{m}_b^2,-\widetilde{x}_e^2) \\\hline  L_{n+d-2}^{(2)}(-\widetilde{x}_c^2) & 
\displaystyle K_{2,n-1}^{(\chi {\rm GUE})}(-\widetilde{x}_c^2,-\widetilde{x}_e^2) \end{array}
\right]_{\tiny\begin{array}{l}  a=1,\ldots,\nu\\b=1,\ldots,N_{\rm f}\\c,e=1,\ldots,k\\d=1,\ldots,N_{\rm f}+\nu \end{array} }\hspace*{-0.2cm}.
\end{eqnarray}
Notice that all factors of $(-1)$ have cancelled, and we recall the definition  \eref{shift-var}. In the same way as we simplified \eref{Zdenom} to \eref{Zdenom.b}, in order to prepare the large-$n$ limit, we can apply here \eref{Lag-recurrence.b} to modify the first $\Nf+\nu$ column as follows:
\begin{eqnarray}
 \fl Z_{2,n-k-1}^{(\tNf)}(\{\widetilde{m}\})&=&\frac{2^{1-n}(n-k-1)!((n-1)!)^{\Nf+\nu}\prod_{j=0}^{n}j!\prod_{l=\nu}^{n-2}l!}{\widetilde{s}^{4\nu}\Delta_{k}(\{x^2\})^2\Delta_{\Nf}(\{m^2\})\left[\prod_{f=1}^{\Nf}m_f^{2\nu}\prod_{j=1}^{k}x_j^{2\nu}(m_f^2+x_j^2)\right]} \nonumber\\
\fl&&\times\det\left[\begin{array}{c|c} 
\partial^{a-1}_{\tilde{s}^2}\widetilde{s}^{2d+2}L_{n-1}^{(d+1)}(-\widetilde{s}^2) & 
\displaystyle \partial_{\tilde{s}^2}^{a-1}\widetilde{s}^4 K_{2,n-1}^{(\chi {\rm GUE})}(-\widetilde{s}^2,-\widetilde{x}_e^2)\\
\hline \widetilde{m}_b^{2d-2}L_{n-1}^{(d+1)}(-\widetilde{m}_b^2) & 
\displaystyle K_{2,n-1}^{(\chi {\rm GUE})}(-\widetilde{m}_b^2,-\widetilde{x}_e^2) \\\hline \widetilde{x}_c^{2d-2} L_{n-1}^{(d+1)}(-\widetilde{x}_c^2) & 
\displaystyle K_{2,n-1}^{(\chi {\rm GUE})}(-\widetilde{x}_c^2,-\widetilde{x}_e^2) \end{array}
\right]_{\tiny\begin{array}{l}  a=1,\ldots,\nu\\b=1,\ldots,N_{\rm f}\\c,e=1,\ldots,k\\d=1,\ldots,N_{\rm f}+\nu \end{array} }\hspace*{-0.2cm}.\quad
\label{Z2prefinal}
\end{eqnarray}
Now, the leading order asymptotic no longer leads to degenerate columns. In a final step, 
we can  carry out the derivatives, where we employ another recurrence relation~\cite[Eq.~(8.971.3)]{Grad}
\begin{equation}\label{Lag-derivative.b}
 (\partial_z-1)z^\nu L_j^{(\nu)}(z)=(j+1)z^{\nu-1}L_{j+1}^{(\nu-1)}(z),
\end{equation}
applied to $z\to-\widetilde{s}^2$. We would like to point out the subtlety that, to be applicable, the power in $z^\nu$ and the order of the generalised Laguerre polynomials $\nu$ have to agree. This is the reason why we multiplied with $\widetilde{s}^4$ in one of the previous steps. After differentiation the power and order are lowered by one unit, which allows to iterate.
Therefore, we recombine the first $\nu$ rows beginning from the highest derivative, so that we obtain shifted derivative operators, $(1+\partial_{\tilde{s}^2})^{a-1}$ instead of $\partial_{\tilde{s}^2}^{a-1}$, making \eref{Lag-derivative.b} in $z\to-\widetilde{s}^2$ applicable. 
We evaluate
\begin{eqnarray}\label{der1}
(1+\partial_{\tilde{s}^2})^{a-1}\widetilde{s}^{2d+2}L_{n-1}^{(d+1)}(-\widetilde{s}^2)&=&\frac{(n+a-2)!}{(n-1)!}\widetilde{s}^{2(d-a+2)}L_{n+a-2}^{(d-a+2)}(-\widetilde{s}^2)
\end{eqnarray}
for the Laguerre polynomials, and for the kernel \eref{Laguerre-kernel} 
\begin{eqnarray}\label{der2}
\left(1+\partial_{\tilde{s}^2}\right)^{a-1} \widetilde{s}^4K_{2,n-1}^{(\chi {\rm GUE})}(-\widetilde{s}^2,-\widetilde{x}_e)
&=&\sum_{j=0}^{n-2} 2\frac{(j+a-1)!}{(j+2)!}\widetilde{s}^{2(3-a)}L_{j+a-1}^{(3-a)}(-\widetilde{s}^2)L_{j}^{(2)}(-\widetilde{x}_e^2)\\
&=&\widetilde{s}^4\frac{(n+a-2)!}{(n-1)!}\widetilde{K}_{2,n-1}^{(a-1)}(-\widetilde{s}^2,-\widetilde{x}_e^2).\nonumber
\end{eqnarray}
In the last line, we have defined 
\begin{equation}
\widetilde{K}_{2,n-1}^{(a-1)}(-\widetilde{s}^2,-\widetilde{x}_e^2)= 
\frac{(n-1)!}{(n+a-2)!}\sum_{j=0}^{n-2} 2\frac{(j+a-1)!}{(j+2)!}\widetilde{s}^{2(1-a)}L_{j+a-1}^{(3-a)}(-\widetilde{s}^2)L_{j}^{(2)}(-\widetilde{x}_e^2),
\label{Tkernel-def}
\end{equation}
which includes the original kernel for $a=1$, $\widetilde{K}_{2,n-1}^{(0)}(-\widetilde{s}^2,-\widetilde{x}_e^2)={K}_{2,n-1}^{(\chi\rm GUE)}(-\widetilde{s}^2,-\widetilde{x}_e^2)$. In such a way, the factors $\widetilde{s}^4$ and factorials in \eref{der1} and \eref{der2} can be pulled out of the first $\nu$ rows of the determinant in~\eref{Z2prefinal}, and we obtain as the final answer for the partition function 
\begin{eqnarray}
 Z_{2,n-k-1}^{(\tNf)}(\{\widetilde{m}\})&=&\frac{2^{1-n}(n-k-1)!((n-1)!)^{\Nf}\prod_{j=0}^{n}j!\prod_{l=0}^{n-2}(l+\nu)!}{\Delta_{k}(\{x^2\})^2\Delta_{\Nf}(\{m^2\})\left[\prod_{f=1}^{\Nf}m_f^{2\nu}\prod_{j=1}^{k}x_j^{2\nu}(m_f^2+x_j^2)\right]} \nonumber\\
&&\times\det\left[\begin{array}{c|c} 
\widetilde{s}^{2(d-a)}L_{n+a-2}^{(d-a+2)}(-\widetilde{s}^2) & 
\displaystyle \widetilde{K}_{2,n-1}^{(a-1)}(-\widetilde{s}^2,-\widetilde{x}_e^2)\\
\hline \widetilde{m}_b^{2d-2}L_{n-1}^{(d+1)}(-\widetilde{m}_b^2) & 
\displaystyle \widetilde{K}_{2,n-1}^{(0)}(-\widetilde{m}_b^2,-\widetilde{x}_e^2) \\\hline \widetilde{x}_c^{2d-2} L_{n-1}^{(d+1)}(-\widetilde{x}_c^2) & 
\displaystyle \widetilde{K}_{2,n-1}^{(0)}(-\widetilde{x}_c^2,-\widetilde{x}_e^2) \end{array}
\right]_{\tiny\begin{array}{l} a=1,\ldots,\nu\\b=1,\ldots,N_{\rm f}\\c,e=1,\ldots,k\\d=1,\ldots,N_{\rm f}+\nu \end{array} }.
\label{Z2final}
\end{eqnarray}

\subsection{Hard edge {\rm k}th spacing distribution at finite-$n$}

Inserting the expression \eref{Z2final} as well as \eref{Zdenom.b} into \eref{level-space.c}, we obtain the following relatively compact expression
for the spacing distribution at the hard edge between the $k$th and $(k+1)$st smallest eigenvalue for arbitrary $k$, with $N_{\rm f}$ flavours and topology $\nu$ at fixed $n$.
Recalling the definition of the shifted variables from \eref{shift-var}, we have
\begin{eqnarray}
 p_{k,n}^{(\nu,\Nf)}(s)&=&\frac{n!\,2}{(k-1)!\,n^{\Nf}(n-1+\nu)!\prod_{f=1}^{\Nf}m_f^{2\nu}\det\left[m_a^{2b-2} L_{n}^{(\nu+b-1)}(-m_a^2)\right]_{a,b=1,\ldots,N_{\rm f}}}\nonumber\\
&&\times \int_{0}^\infty dx_{k}\int_0^{x_{k}}dx_1
\ldots \int_0^{x_{k}}dx_{k-1} \widetilde{s}^{2\nu+1}e^{-(n-k)\widetilde{s}^2}\prod_{f=1}^{\Nf}\widetilde{m}_f^{2}\prod_{j=1}^{k}\widetilde{x}_j^4x_je^{-x_j^2}\nonumber\\
&&\times\det\left[\begin{array}{c|c} 
\widetilde{s}^{2(d-a)}L_{n+a-2}^{(d-a+2)}(-\widetilde{s}^2) & 
\displaystyle \widetilde{K}_{2,n-1}^{(a-1)}(-\widetilde{s}^2,-\widetilde{x}_e^2)\\
\hline \widetilde{m}_b^{2d-2}L_{n-1}^{(d+1)}(-\widetilde{m}_b^2) & 
\displaystyle \widetilde{K}_{2,n-1}^{(0)}(-\widetilde{m}_b^2,-\widetilde{x}_e^2) \\\hline \widetilde{x}_c^{2d-2} L_{n-1}^{(d+1)}(-\widetilde{x}_c^2) & 
\displaystyle \widetilde{K}_{2,n-1}^{(0)}(-\widetilde{x}_c^2,-\widetilde{x}_e^2) \end{array}
\right]_{\tiny\begin{array}{l}  a=1,\ldots,\nu\\b=1,\ldots,N_{\rm f}\\c,e=1,\ldots,k\\d=1,\ldots,N_{\rm f}+\nu \end{array} }.
\label{level-space.e}
\end{eqnarray}
The expression~\eref{level-space.e} is our first main result and will serve as the starting point for the asymptotic analysis in the next section.

	\begin{figure}[t!]
		\centerline{
		\includegraphics[width=0.49\textwidth]{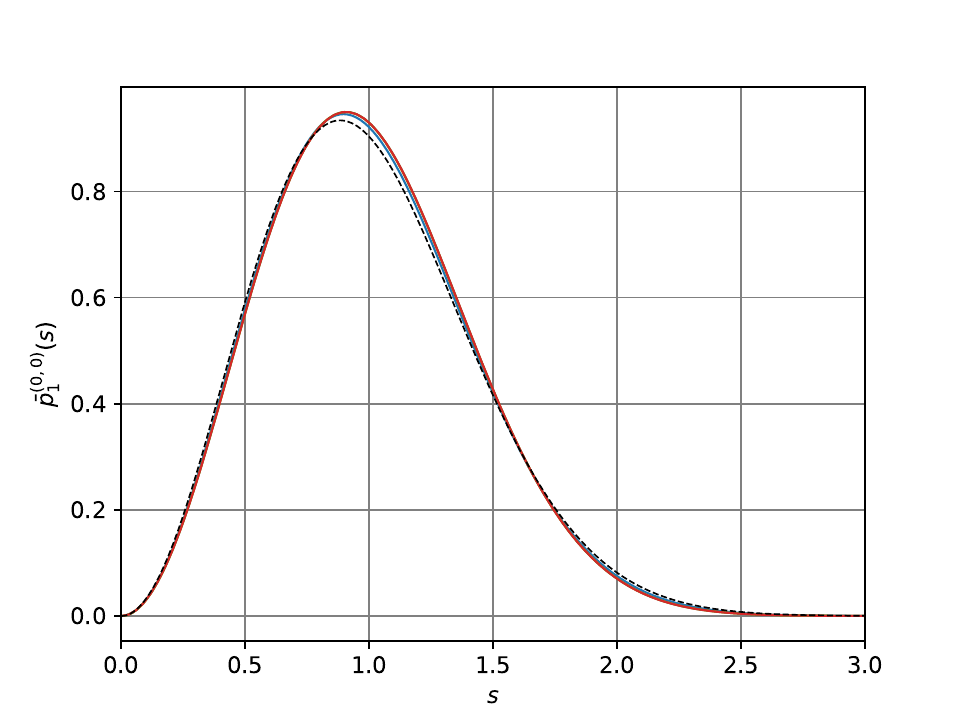}	\includegraphics[width=0.49\textwidth]{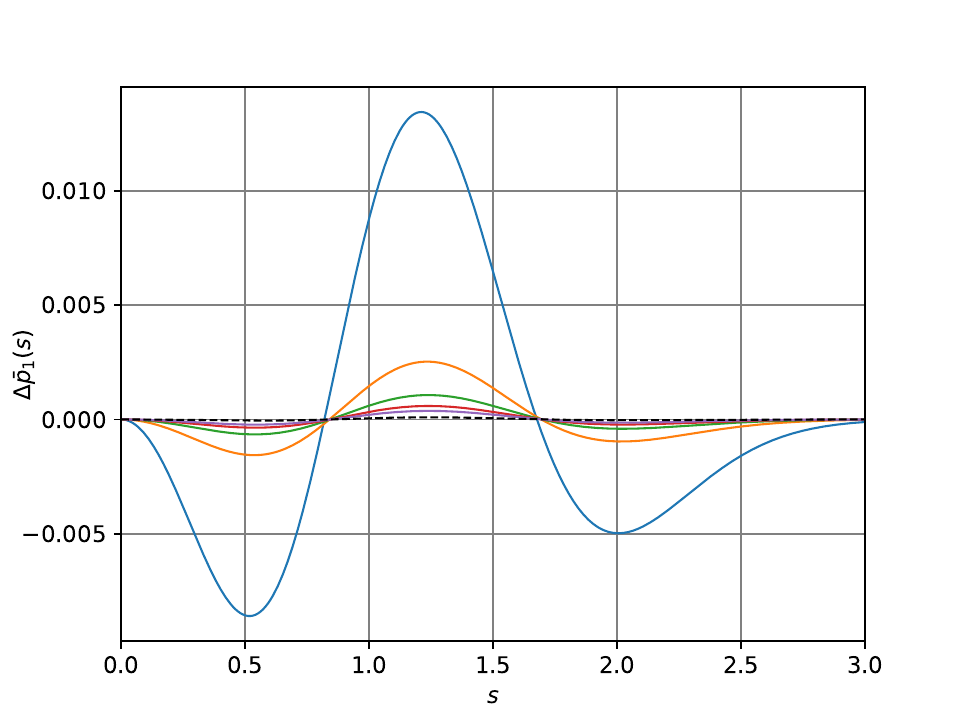}
		}
		\centerline{
		\includegraphics[width=0.49\textwidth]{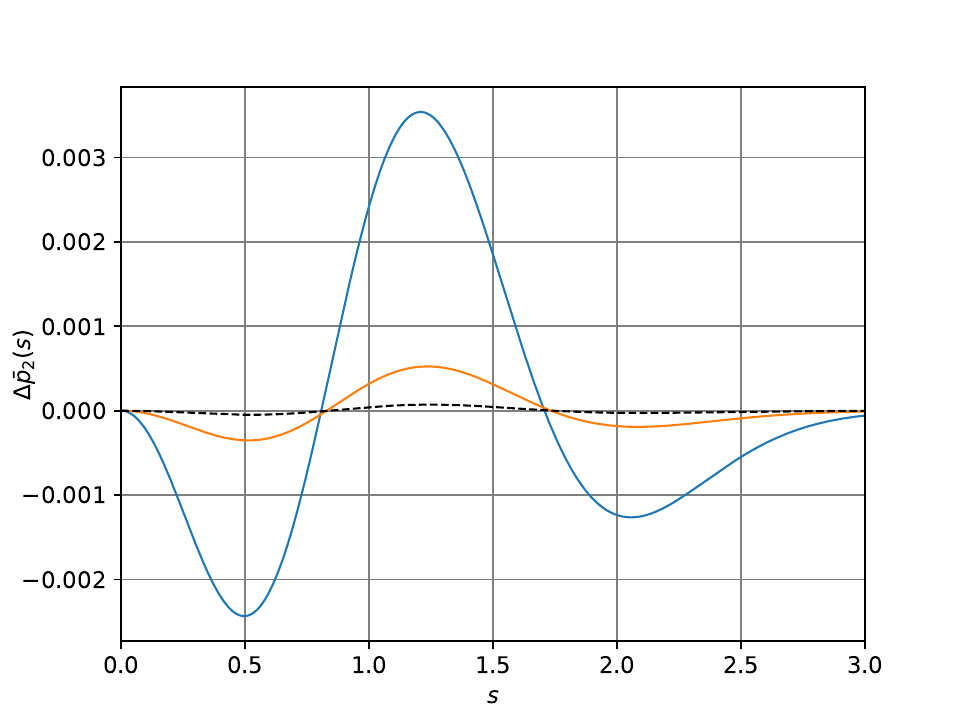}
		\includegraphics[width=0.49\textwidth]{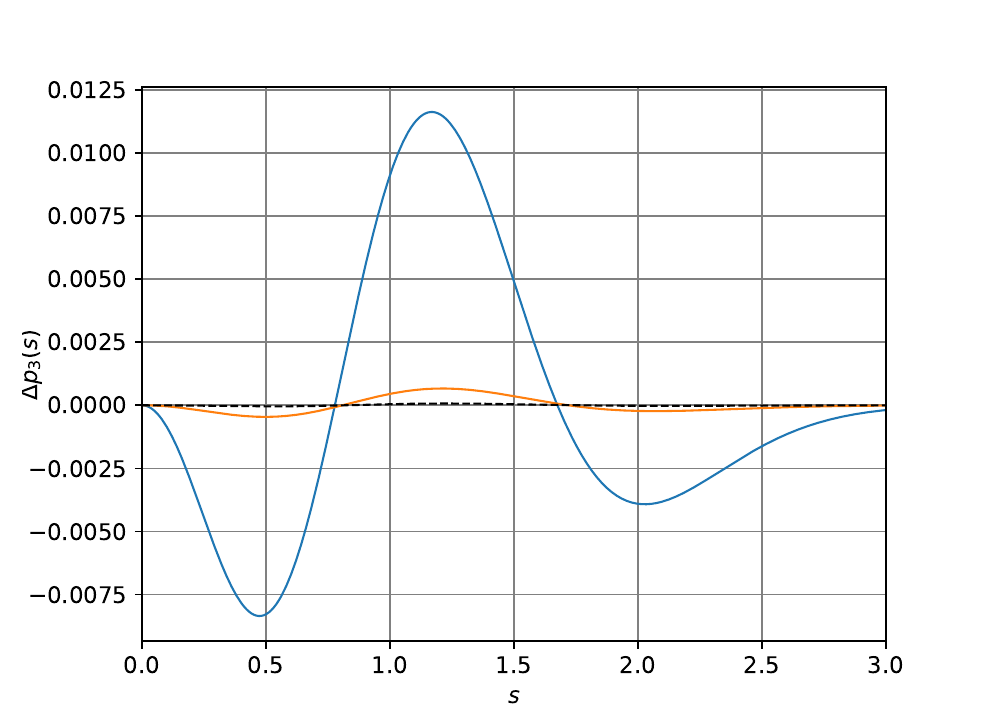}
		}
		\caption{{\bf Top Left:} The spacing distribution $\bar{p}_{1,n}^{(0,0)}(\sigma)$  between 1st and 2nd eigenvalue ($k=1$) at the hard edge, rescaled to first moment of unity, for different values of $n=2\,{\rm (blue)}$, $n=6\,{\rm (orange)}$, $n=10\,{\rm (green)}$, and $n= 20\,{\rm (red)}$. The dashed curve is the bulk spacing distribution~\eref{eq:bulk-spacing} for comparison. {\bf Top Right:} Because the curves are almost indistinguishable we also compare the difference $\Delta p_k
		(s)$ between the asymptotic spacing~\eref{p2rescaled} to be derived in the next section, and our expression at finite-$n$ \eref{level-den-finite.b}, for $n=2\,{\rm (blue)}$, $n=4\,{\rm (orange)}$, $n=6\,{\rm (green)}$, $n=8\,{\rm (red)}$, $n=10\,{\rm (purple)}$, and $n=20\,{\rm (dashed black)}$.
		\newline  {\bf Bottom Row:} Differences of the finite $n$ spacing~\eref{level-space.e} to their asymptotic counterpart~\eref{lim-level-space} for $k=2,\, n=4, 6, 8, 10, 20$ ({\bf Left}), and $k=3,\, n=4, 8, 20$ ({\bf Right}). In both plots the dashed line denotes the difference for $n=20$, both with their first moment set to unity. These plots illustrate the rapid convergence already at relatively small values of $n$. Please note the different scales of the $y$-axis.
		 }
		\label{Fig:Bulk-p2N2-20}
	\end{figure}

Before we come to this, let us make some comments on the duality between the index $\nu$ and the number of flavours with mass $0$. Indeed, when taking the limit $m_{N_{\rm f}}\to0$, the result~\eref{level-space.e} reduces to the case $N_{\rm f}\to N_{\rm f}-1$ and $\nu\to\nu+1$. This can be readily checked, while the other limit $m_{N_{\rm f}}\to\infty$ leads to  the partially quenched case $N_{\rm f}\to N_{\rm f}-1$ and $\nu\to\nu$, meaning the index $\nu$ stays the same.

Let us give some simple examples. In the absence of flavours ($\Nf=0$) and zero-modes ($\nu=0$), we obtain  the following expression for the $k$th spacing in the $\chi$GUE with square matrices
\begin{eqnarray}\label{level-space-quench-k}
 p_{k,n}^{(0,0)}(s)&=&\frac{2n}{(k-1)!}\int_{0}^\infty dx_{k}\int_0^{x_{k}}dx_1
\ldots \int_0^{x_{k}}dx_{k-1} \prod_{j=1}^{k}((x_{k}+s)^2-x_j^2)^2x_je^{-x_j^2}
\\
&&\times
(x_{k}+s)\,e^{-(n-k)(x_{k}+s)^2}
\det\left[ {K}_{2,n-1}^{(\chi\rm GUE)}({x}_c^2-(x_{k}+s)^2,{x}_e^2-(x_{k}+s)^2) 
\right]_{c,e=1}^{k}.
\nonumber
\end{eqnarray}
Here, all shifted variables are explicitly written out. In particular, for the spacing between the first and second eigenvalue ($k=1$) we obtain a one-fold integral representation
\begin{eqnarray}  \label{level-space-quench}
\!\!\!\!\!p_{1,n}^{(0,0)}(s)&=&4\int_{0}^\infty dx_{1}s^2(2x_{1}+s)^2x_1(x_{1}+s)e^{-x_1^2-(n-1)(x_{1}+s)^2}
\\
&&\times\left(L_{n-1}^{(2)}(-s(2x_{1}+s))L_{n-2}^{(2)}(-s(2x_{1}+s))-L_{n-2}^{(3)}(-s(2x_{1}+s))L_{n-1}^{(1)}(-s(2x_{1}+s))\right),
\nonumber
\end{eqnarray}
where we inserted~\eref{level-den-finite.b} for the kernel at equal arguments. 
This result still has to be unfolded, as explained at the end of Section~\ref{sec:finiteN}.

In Figure~\ref{Fig:Bulk-p2N2-20} (top row), we compare~\eref{level-space-quench} for small values of $n$ and the convergence to the asymptotic limit. The result for $k=1$ at $n=2$ is already very close to the limiting result, showing the optimal rate of convergence $1/n^2$ at the hard edge, as it was shown in~\cite{ForresterTrinh}. This certainly is also true when increasing $k$ to $k=2$ and $k=3$, see bottom row of Figure~\ref{Fig:Bulk-p2N2-20}. The reason why the optimal rate of convergence applies here is that we unfold via the formula~\eref{unfolded-level-density}, since in~\cite{ForresterTrinh} it was shown that already a scaling is sufficient to reach this rate.

Since the deviations of the finite $n$-results and the asymptotic ones lie in the per mill regime, we have plotted the difference $\Delta p_k(s)=\bar{p}_{k,n}^{(0,0)}(s)-\bar{p}_k^{(0,0)}(s)$. The bar highlights the unfolding~\eref{unfolded-level-density}, and $p_k^{(0,0)}(s)$ is the asymptotic result~\eref{lim-level-space}.

\section{Asymptotic Analysis at the Hard Edge}\label{sec:largeN}
\subsection{The large-$n$ limit at fixed $k$, $N_{\rm f}$ and $\nu$}

In the asymptotic large-$n$ limit of the $\chi$GUE at the hard edge, with our convention of an $n$-independent weight function,  it is well known that the eigenvalues, and therefore also the spacings and masses scale with $\sqrt{n}$, see~\cite{reviewQCD}. We therefore define the following variables
\begin{equation}
\sigma\equiv\sqrt{n}\,s,\quad \mu_f\equiv\sqrt{n}\, m_f, \quad f=1,\ldots,\Nf,
\label{scaling1}
\end{equation}
where $\sigma$ and the $\mu_f$ remain finite in the large-$n$ limit. In view of the integrations to be performed in~\eref{level-space.e}, and the definition~\eref{shift-var} of the shifted variables therein, we also redefine the integration variables in the following fashion:
\begin{equation}
\chi\equiv \sqrt{n}\, x_{k}\quad{\rm and}\quad 
\lambda_j\equiv\sqrt{n}\,x_j,  \quad j=1,\ldots,k-1.
\label{scaling2}
\end{equation}
We begin with the large-$n$ asymptotic of the generalised Laguerre polynomials, using~\cite[Eq.~(18.15.19)]{NIST}
 \begin{equation}\label{Laguerre-asymp}
 \lim_{\small\begin{array}{c} n,j\to\infty\\j/n\to t \end{array}}\frac{1}{n^\nu} L_j^{(\nu)}\left(-\frac{z^2}{n}\right)=\left(\frac{\sqrt{t}}{z}\right)^\nu I_{\nu}\left(2z\sqrt{t}\right),
 \end{equation}
 with $I_\nu(z)$ the modified Bessel function of the first kind. In foresight, we already allow here for a general degree $j$ of the polynomial, to be able to take the limit under the sum in \eref{Tkernel-def}. When $j=\mathcal{O}(n)$ we can set $t=1$ in the above expression, as it will be the case for the unintegrated Laguerre polynomials. We apply this first to the mass dependent determinant of Laguerre polynomials in the denominator in the first line of~\eref{level-space.e}. In view of the scaling~\eref{scaling1}, we also include the mass dependent pre-factor $\prod_{f=1}^{\Nf}m_f^{2\nu}$ into the determinant, to obtain
 \begin{equation}
\lim_{n\to\infty} \det\left[(\mu_a^2/n)^{\nu+b-1} L_{n}^{(\nu+b-1)}(-\mu_a^2/n)\right]_{a,b=1,\ldots,N_{\rm f}}
 =\det\left[\mu_a^{\nu+b-1} I_{\nu+b-1}(2\mu_a)\right]_{a,b=1,\ldots,N_{\rm f}}.
 \end{equation}
 
Next, we move to the building blocks of the large determinant in the last line of~\eref{level-space.e}. We begin with distinguishing the Laguerre polynomials of different variables, multiplied by different powers. Inserting Eqs.~\eref{scaling1} and~\eref{scaling2}, we obtain from the limit~\eref{Laguerre-asymp}
\begin{eqnarray}
\widetilde{s}^{2(d-a)}L_{n+a-2}^{(d-a+2)}(-\widetilde{s}^2) & =&
\left(\frac{\chi+\sigma}{\sqrt{n}}\right)^{2(d-a)}L_{n+a-2}^{(d-a+2)}\left(-\frac{(\chi+\sigma)^2}{n}\right) 
\nonumber\\
&\approx&n^2(\chi+\sigma)^{d-a-2}I_{d-a+2}(2(\chi+\sigma)),
\nonumber\\
\widetilde{m}_b^{2d-2}L_{n-1}^{(d+1)}(-\widetilde{m}_b^2) & \approx&n^2
[(\chi+\sigma)^2+\mu_b^2]^{(d-3)/2}I_{d+1}\left(2\sqrt{(\chi+\sigma)^2+\mu_b^2}\right), \quad b=1,\ldots,\Nf,\nonumber\\
\widetilde{x}_c^{2d-2}L_{n-1}^{(d+1)}(-\widetilde{x}_c^2) & \approx&n^2
[(\chi+\sigma)^2-\lambda_c^2]^{(d-3)/2}I_{d+1}\left(2\sqrt{(\chi+\sigma)^2-\lambda_c^2}\right), \quad c=1,\ldots,k-1,\nonumber\\
\widetilde{x}_{c}^{2d-2}L_{n-1}^{(d+1)}(-\widetilde{x}_{c}^2) & \approx&n^2[\sigma(2\chi+\sigma)]^{(d-3)/2}I_{d+1}\left(2\sqrt{\sigma(2\chi+\sigma)}\right),\quad c=k.
\label{Laguerre-all-asymp}
\end{eqnarray}
In view of these results, it is convenient to define again a set of limiting shifted variables, in analogy to~\eref{shift-var}, in order to compactify the final answer for the limiting spacing distribution:
\begin{eqnarray}
&&\widetilde{\sigma}\equiv \chi+\sigma=\sqrt{n}\,\widetilde{s},\quad\quad\quad\quad\quad\,\widetilde{\mu}_f\equiv \sqrt{(\chi+\sigma)^2+\mu_f^2}=\sqrt{n}\,\widetilde{m}_f, \quad f=1,\ldots,\Nf,\nonumber\\
&&\widetilde{\chi}\equiv \sqrt{\sigma(2\chi+\sigma)}=\sqrt{n}\,\widetilde{x}_{k},\quad \widetilde{\lambda}_c\equiv \sqrt{(\chi+\sigma)^2-\lambda_c^2}=\sqrt{n}\,\widetilde{x}_c, \quad c=1,\ldots,k-1.\quad 
\label{shift-large}
\end{eqnarray}

We are now ready to take the limit of the kernel~ \eref{Tkernel-def}. For any $e=1,\ldots,k-1$, it holds
\begin{eqnarray}
\widetilde{K}_{2,n-1}^{(a-1)}(-\widetilde{s}^2,-\widetilde{x}_e^2)&\approx&
\frac{2}{n^{a-1}}\sum_{j=0}^{n-2}j^{a-3}\left(\frac{\widetilde{\sigma}^2}{n}\right)^{1-a}L_{j+a-1}^{(3-a)}\left(-\frac{\widetilde{\sigma}^2}{n}\right)L_j^{(2)}
\left(-\frac{\widetilde{\lambda}^2_e}{n}\right)\nonumber\\
&\approx& 2n^3\int_0^1dt \,t^{\frac{a-1}{2}}\widetilde{\sigma}^{-1-a}I_{3-a}\left(2\widetilde{\sigma}\sqrt{t}\right)\widetilde{\lambda}_e^{-2}I_{2}\left(2\widetilde{\lambda}_e\sqrt{t}\right)\nonumber\\
&
=& n^3 \mathcal{K}_{a-1}\left(\widetilde{\sigma},\widetilde{\lambda}_e\right), 
\end{eqnarray}
where we have defined 
\begin{equation}
\mathcal{K}_{a-1}\left(\widetilde{\sigma},\widetilde{\lambda}_e\right)\equiv
4\widetilde{\sigma}^{-1-a}\widetilde{\lambda}_e^{-2}\int_0^1dT\, T^aI_{3-a}\left(2\widetilde{\sigma}_{}T\right)I_{2}(2\widetilde{\lambda}_eT),
\label{Ka-1def}
\end{equation}
 after changing variables. The same result holds for $e=k$, replacing $\widetilde{\lambda}_e$ by $\widetilde{\chi}$. Finally, for $a=1$ we obtain for the limit of the $\chi$GUE kernel:
\begin{eqnarray}
\widetilde{K}_{2,n-1}^{(0)}(-\widetilde{m}_b^2,-\widetilde{x}_e^2)&\approx&n^3
\mathcal{K}_{0}\left(\widetilde{\mu}_b,\widetilde{\lambda}_e\right)=n^3
4\widetilde{\mu}_b^{-2}\widetilde{\lambda}_e^{-2}\int_0^1dT\, TI_{2}\left(2\widetilde{\mu}_{b}T\right)I_{2}(2\widetilde{\lambda}_eT)\
\nonumber\\
&=& \frac{4n^3}{(\widetilde{\mu}_b\widetilde{\lambda}_e)^2(\widetilde{\lambda}_e^2-\widetilde{\mu}_b^2)}\left( 2\widetilde{\lambda}_eI_1(2\widetilde{\lambda}_e)I_2(2\widetilde{\mu}_b)-2\widetilde{\mu}_bI_1(2\widetilde{\mu}_b)I_2(2\widetilde{\lambda}_e)\right) ,
\end{eqnarray}
 for $b=1,\dots,\Nf$ and $e=1,\ldots,k-1$, and likewise for different arguments 
$\widetilde{\mu}_b\to\widetilde{\lambda}_c$ and $\widetilde{\lambda}_e\to\widetilde{\chi}$. In the last step we have performed the integral, corresponding to the limit  of the second line in \eref{Laguerre-kernel}, whereas the first line follows from replacing the sum with an integral. Finally, for equal arguments we have 
\begin{equation}
\label{limKxx}
\widetilde{K}_{2,n-1}^{(0)}(-\widetilde{x}_c^2,-\widetilde{x}_c^2)\approx n^3
\mathcal{K}_{0}\left(\widetilde{\lambda}_c,\widetilde{\lambda}_c\right)=n^3
\frac{2}{\widetilde{\lambda}_c^4}\left(I_2(2\widetilde{\lambda}_c)^2-I_3(2\widetilde{\lambda}_c)I_1(2\widetilde{\lambda}_c)\right),
\end{equation}
and likewise for $\widetilde{\lambda}_c\to\widetilde{\chi}$. If we had taken the limit for Laguerre polynomials with positive argument in \eref{Laguerre-asymp}, this would be proportional to the Bessel density at $\nu=2$, see \eref{Bessel-density}. The factor of two in the argument of the Bessel functions indicates that in our scaling limit~\eref{limKxx} the mean level spacing is $\pi/2$. We will come back to this below.

We now have all ingredients together to take the limit of the spacing distribution \eref{level-space.e}, where in view of the scaling \eref{scaling1} we define
\begin{eqnarray}
\label{lim-level-space}
\fl p_k^{(\nu,\Nf)}(\sigma)&\equiv&\lim_{n\to\infty}\frac{1}{\sqrt{n}}  p_{k,n}^{(\nu,\Nf)}\left( \frac{s}{\sqrt{n}}\right)
\\
\fl&=& \frac{2}{(k-1)!\det\left[\mu_a^{\nu+b-1} I_{(\nu+b-1)}(2\mu_a)\right]_{a,b=1,\ldots,N_{\rm f}}}\int_0^\infty d\chi\int_0^\chi d\lambda_1\cdots \int_0^\chi d\lambda_{k-1}\widetilde{\sigma}^{2\nu+1}e^{-\tilde{\sigma}^2}\widetilde{\chi}^4\chi\nonumber\\
\fl&&\times\prod_{f=1}^{\Nf}\widetilde{\mu}_f^2\left(\prod_{j=1}^{k-1}\widetilde{\lambda}_j^4\lambda_j\right)\ \det\left[\begin{array}{c|c|c} 
\widetilde{\sigma}^{d-a-2}I_{d-a+2}(2\widetilde{\sigma}) & 
\displaystyle \mathcal{K}_{a-1}\left(\widetilde{\sigma},\widetilde{\lambda}_e\right)& 
\displaystyle \mathcal{K}_{a-1}\left(\widetilde{\sigma},\widetilde{\chi}\right)
\\
\hline \widetilde{\mu}_b^{d-3}I_{d+1}(2\widetilde{\mu}_b) & 
\displaystyle \mathcal{K}_{0}\left(\widetilde{\mu}_b,\widetilde{\lambda}_e\right)  & 
\displaystyle \mathcal{K}_{0}\left(\widetilde{\mu}_b,\widetilde{\chi}\right) \\
\hline \widetilde{\lambda}_c^{d-3} I_{d+1}(2\widetilde{\lambda}_c) & 
\displaystyle \mathcal{K}_{0}\left(\widetilde{\lambda}_c,\widetilde{\lambda}_e\right) & 
\displaystyle \mathcal{K}_{0}\left(\widetilde{\lambda}_c,\widetilde{\chi}\right) \\
\hline \widetilde{\chi}^{d-3} I_{d+1}(2\widetilde{\chi}) & 
\displaystyle \mathcal{K}_{0}\left(\widetilde{\chi},\widetilde{\lambda}_e\right) & 
\displaystyle \mathcal{K}_{0}\left(\widetilde{\chi},\widetilde{\chi}\right) \\
\end{array}
\right]_{\tiny\begin{array}{l} a=1,\ldots,\nu\\b=1,\ldots,N_{\rm f}\\c,e=1,\ldots,k-1\\d=1,\ldots,N_{\rm f}+\nu \end{array} }.\nonumber
\end{eqnarray}
In the quenched case $\Nf=0$ of the $\chi$GUE with $\nu=0$  it reduces to
\begin{eqnarray}
\label{lim-level-space00}
\fl p_k^{(0,0)}(\sigma)
= \frac{2}{(k-1)!}\int_0^\infty d\chi\int_0^\chi d\lambda_1\cdots \int_0^\chi d\lambda_{k-1}\widetilde{\sigma}e^{-\tilde{\sigma}^2}\widetilde{\chi}^4\chi\prod_{j=1}^{k-1}\widetilde{\lambda}_j^4\lambda_j
\det\left[\begin{array}{c|c} 
\displaystyle \mathcal{K}_{0}\left(\widetilde{\lambda}_c,\widetilde{\lambda}_e\right) & 
\displaystyle \mathcal{K}_{0}\left(\widetilde{\lambda}_c,\widetilde{\chi}\right) \\
\hline
\displaystyle \mathcal{K}_{0}\left(\widetilde{\chi},\widetilde{\lambda}_e\right) & 
\displaystyle \mathcal{K}_{0}\left(\widetilde{\chi},\widetilde{\chi}\right) \\
\end{array}
\right]_{c,e=1,\ldots,k-1}\hspace*{-1cm}.
\end{eqnarray}
In the simplest case $k=1$ (spacing between the first and second smallest eigenvalue), we thus obtain the following single integral representation
\begin{eqnarray}
\label{lim-level-space00k2}
\fl p_1^{(0,0)}(\sigma)
= 4\int_0^\infty d\chi\chi(\chi+\sigma)e^{-(\chi+\sigma)^2}
\left(I_2\left(2\sqrt{\sigma(2\chi+\sigma)}\right)^2-I_3\left(2\sqrt{\sigma(2\chi+\sigma)}\right)I_1\left(2\sqrt{\sigma(2\chi+\sigma)}\right)\right),
\end{eqnarray}
where we have spelled out  all shifted variables explicitly, as well as the kernel from \eref{limKxx}.

	\begin{figure}[t!]
		\centerline{
		\includegraphics[width=0.49\linewidth,angle=0]{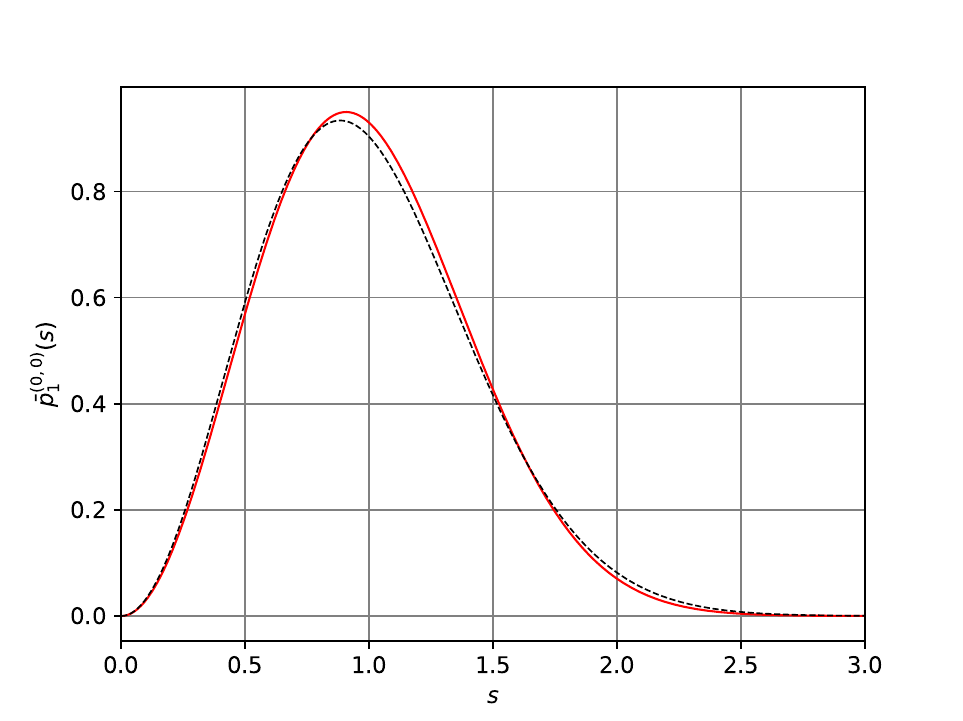}
		\includegraphics[width=0.49\linewidth,angle=0]{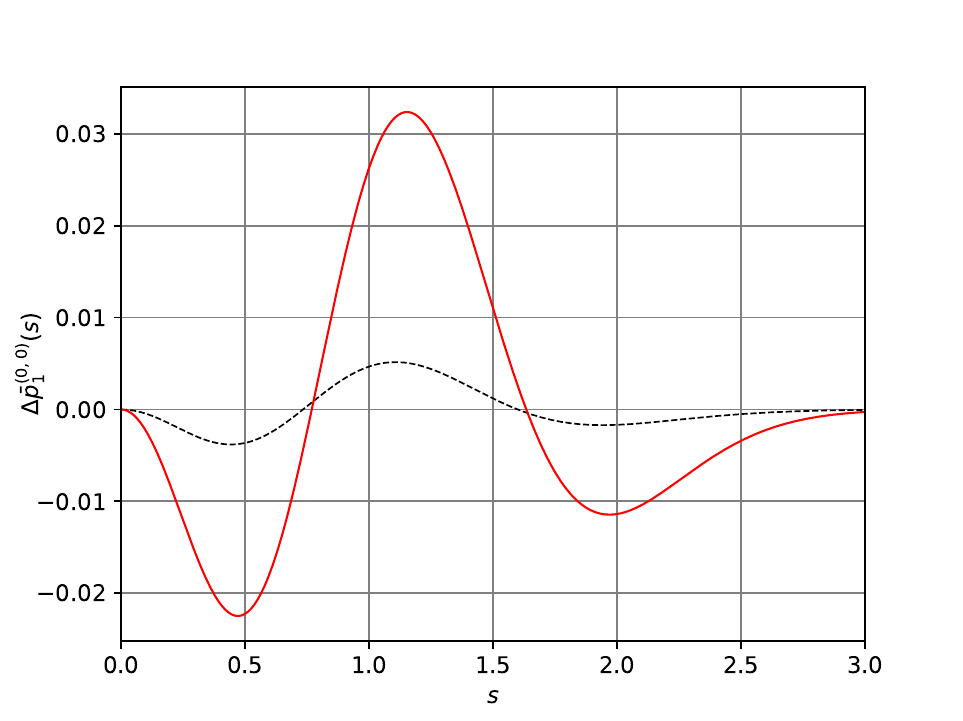}}
		\caption{{\bf Left:} Comparison between the bulk spacing distributions $p_{\rm bulk}(s)$ (see~\eref{eq:bulk-spacing}) (dashed black, lower curve), and the spacing distribution between 1st and 2nd eigenvalue at the hard edge $\bar{p}_1(s)$ with $k=1$ and $N_{\rm f}=\nu=0$ quenched (red, higher curve), see~\eref{lim-level-space00k2} and \eref{p2rescaled}. {\bf Right:} Their difference $\Delta p_1(s)=\bar  p_1^{(0,0)}(s)-p_{\rm bulk}(s)$ (red higher curve) is still considerably larger than the difference between the exact bulk spacing and the Wigner surmise (black, dashed curve).
		}
		\label{Fig:Bulk-p2}
	\end{figure}

As explained at the end of Section \ref{sec:gap}, we still have to unfold, by computing the first moment and rescaling the spacing distribution accordingly. Here and in the setting with more parameters, we were only able to do this numerically. In the present case this means
\begin{equation}
\bar{\sigma}_1=\int_0^\infty d\sigma\sigma p_1^{(0,0)}(\sigma) \approx 1.509,
\end{equation}
leading to the final answer 
\begin{equation}
\label{p2rescaled}
\bar{p}_1^{(0,0)}(\sigma)= \bar{\sigma}_1p_1^{(0,0)}(\bar{\sigma}_1\sigma).
\end{equation}
As it was argued after~\eref{limKxx}, the approximate mean level spacing we expect here is $\pi/2\approx 1.571$. It is very close to $\bar{\sigma}_1$, but given the very small deviation from the bulk spacing to be discussed below, we better use its exact value $\bar{\sigma}_1$ to set the first moment of $\bar{p}_1^{(0,0)}(\sigma)$ exactly to unity.

\subsection{Comparison to the bulk spacing distribution}\label{sec:comp}

We now turn to the comparison between the hard edge spacing distribution for different parameter values $k$, $N_{\rm f}$ and $\nu$, and the bulk spacing distribution 
\eref{eq:bulk-spacing}. We will also compare with the Wigner surmise~\eref{Wignersurmise}, which is often used in comparison to real data, in order to illustrate the closeness between the hard edge and bulk spacing.

	\begin{figure}[t!]
		\centerline{
		\includegraphics[width=0.49\textwidth]{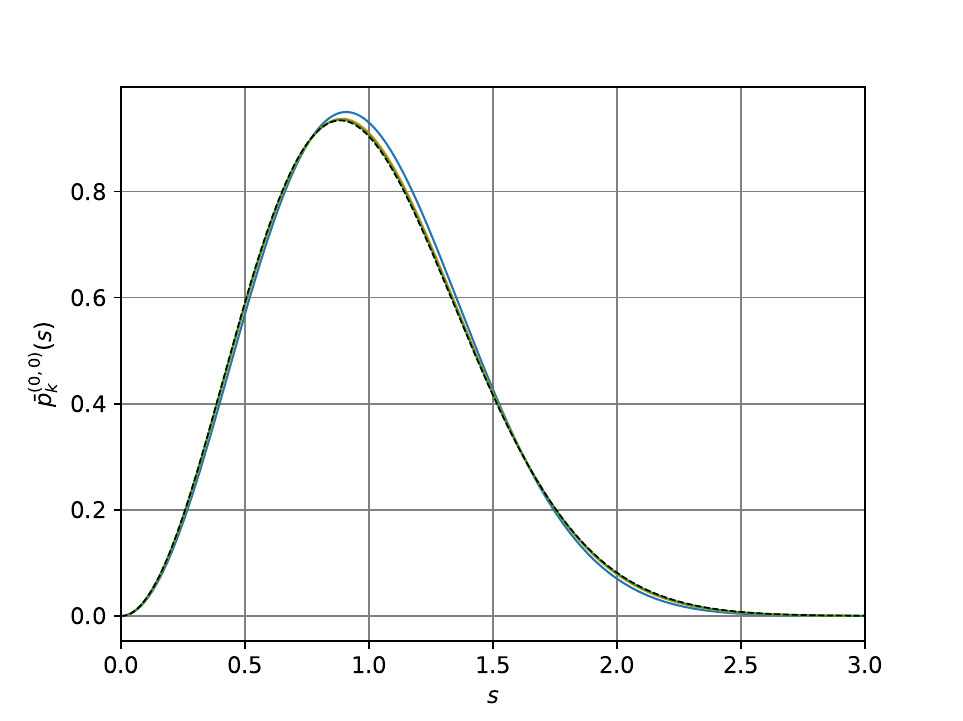}	\includegraphics[width=0.49\textwidth]{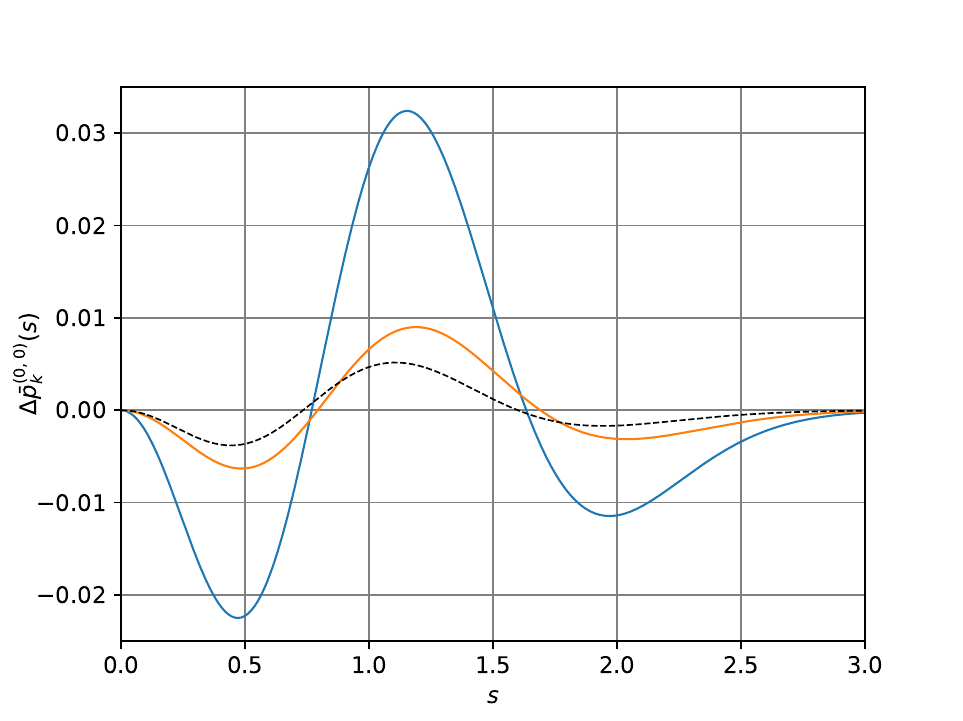}
		}
		\centerline{
		\includegraphics[width=0.49\textwidth]{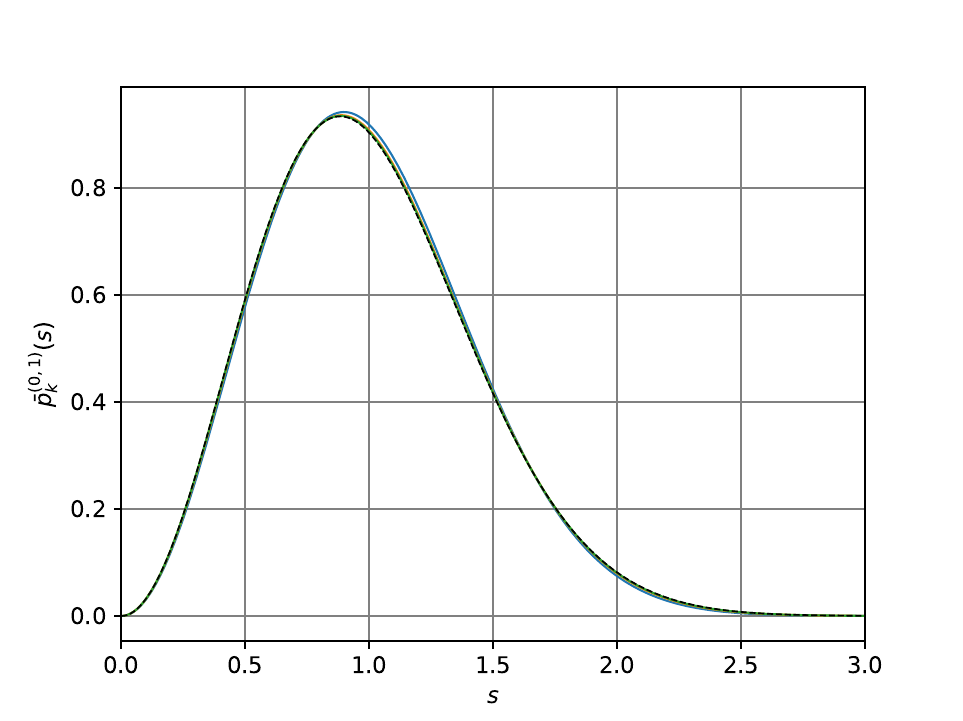}	\includegraphics[width=0.49\textwidth]{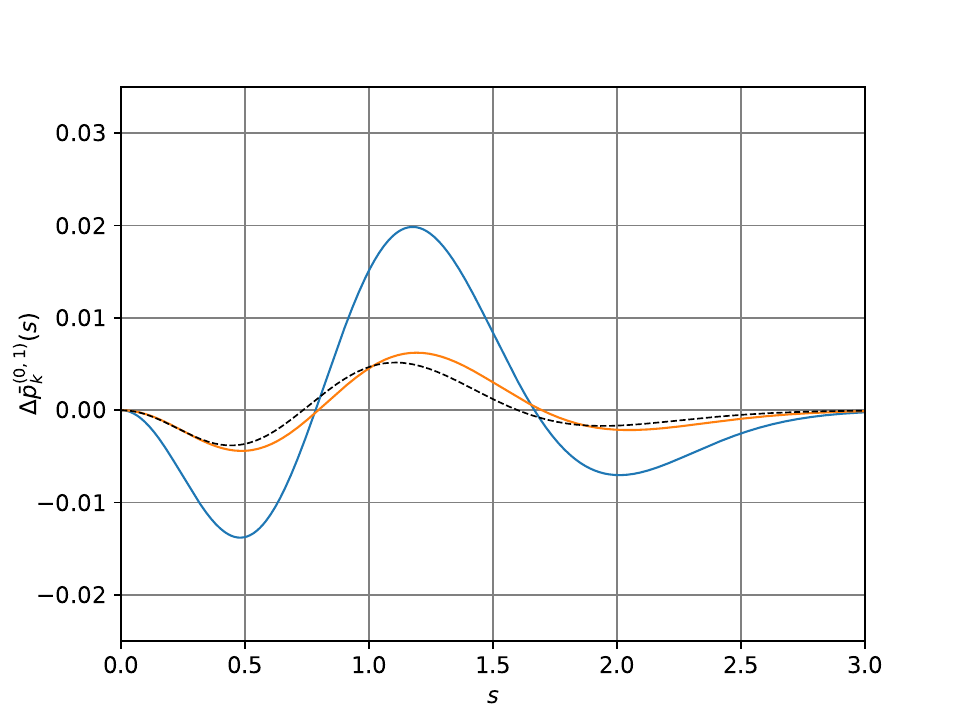}
		}
		\caption{{\bf Left:} Comparison between hard edge level spacings for increasing $k=1\,{\rm (blue), }$ $k=2\,{\rm (orange),\ and }\ k=3\,{\rm (green)}$, and the bulk spacing distribution for $N_{\rm f}=0$ (upper plots), and $N_{\rm f}=1$ (lower plots) with mass $m=0.1$. {\bf Right:}
Their differences $\Delta p_k^{(0,\Nf)}(s)=\bar  p_k^{(0,\Nf)}(s)-p_{\rm bulk}(s)$ to the bulk spacing distribution~\eref{eq:bulk-spacing} for $k=1\,{\rm (blue) }$ and $k=2\,{\rm (orange)}$ decrease rapidly for increasing $k$, where for comparison we also show the difference between the exact bulk spacing and the Wigner surmise (black, dashed curve).
}\label{fig:Comp_As-Bulk}
	\end{figure}	

We begin with the quenched spacing between the first two smallest eigenvalues ($k=1$) with proper unfolding, meaning the first moment is unity, see Figure~\ref{Fig:Bulk-p2}. The deviations of this level spacing distribution from the Wigner surmise~\eref{Wignersurmise} or from the bulk spacing distribution~\eref{eq:bulk-spacing} is about one percent and can be discerned with the bare eye at the maximum of the distribution. This has to be seen in contrast to the difference between the Wigner surmise and the bulk spacing distribution which is only about a few per mill, see~\cite{DH}. As already mentioned at the end of Section~\ref{sec:heuristic}, the case $k=1$ and $\nu=N_{\rm f}=0$ will be the one that should show the strongest deviations.

One approach to the bulk level spacing is certainly given when increasing $k$, as we literally move into the bulk. For simplicity we restrict ourselves to the quenched case ($N_{\rm f}=0$) and a single flavour ($N_{\rm f}=1$), see Figure~\ref{fig:Comp_As-Bulk}. We find, that the deviation from the bulk spacing decreases very quickly. Already for $k=3$ the difference to the bulk spacing distribution is of the same order as the difference between the bulk level spacing distribution~\eref{eq:bulk-spacing} and Wigner's surmise~\eref{Wignersurmise}. This agrees with our observation for the microscopic level density at the end of Section~\ref{sec:heuristic}. Introducing flavours only suppresses this difference even more.

A warning is in order. The numerical evaluation of the multiple  $k$-fold integral representation~\eref{lim-level-space00} becomes rapidly unstable when going to higher values of $k$, $\nu$ and $N_{\rm f}$. One reason is the increasing number of integrals to be carried out. Another reason is the ratio of the determinants in the integrand, which can become numerically unstable as the size of the determinant increases. It may happen that  big as well as tiny numbers have to  cancel. Metropolis-algorithms are also hard to implement, though they are in principle possible, because the functions in the integrand are quite involved.

 To overcome this problem we have employed Monte-Carlo simulations of the random matrix ensemble to study the change of the level spacing distribution when increasing $k$, in particular for $k\geq4$. For this purpose, we only consider parameter sets with $N_{\rm f}=0$, since otherwise we would need to rely on the Metropolis-Hastings algorithm. This again would exceed feasible computation time for a sufficiently large matrix dimension and number of configurations. Furthermore, due to the duality between $N_{\rm f}$ and $\nu$, it is equivalent to only consider $\nu\neq0$ or $N_{\rm f}\neq0$ with masses very close to zero, as discussed in Section~\ref{sec:finiteN}.
	
		As the eigenvalue statistics at the hard and the soft edge are fairly different, as can be seen in Figure~\ref{fig:Bessel-Airy-trans}, where the microscopic spectral densities of those two regions of the spectrum are displayed, we further want to study the level spacings obtained from  numerical simulations, first for relatively small $k$, and secondly for the transition from the bulk to the soft edge. To get the clearest signal, we have first considered $\nu=0$, as for larger $\nu$ we are closer to the soft edge, cf., Figure~\ref{fig:Bessel-Airy-trans}.
		
		\begin{figure}[t!]
		\begin{tikzpicture}[scale=1]
			\draw (-3,1.9) node{\includegraphics[width=0.5\textwidth]{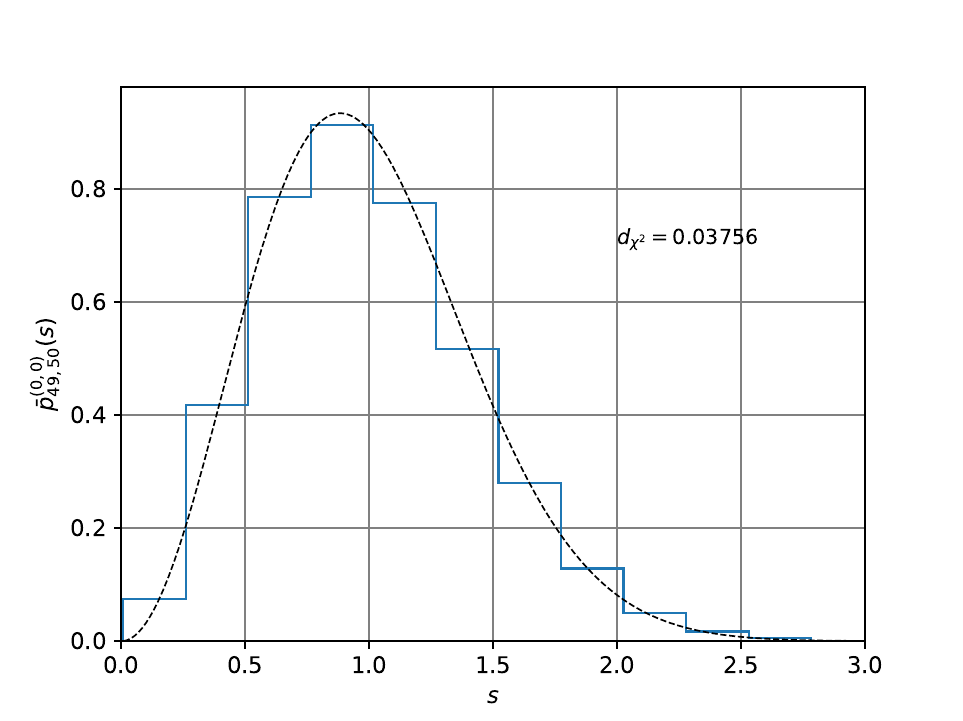}};
			\draw (-1.,2.9) node[black,fill=white,rounded corners=2pt,inner sep=1pt,opacity=1,text opacity=1]{$d_{\chi^2}=0.20144$};
		\end{tikzpicture}
		\begin{tikzpicture}[scale=1]
			\draw (-3,1.9) node{\includegraphics[width=0.5\textwidth]{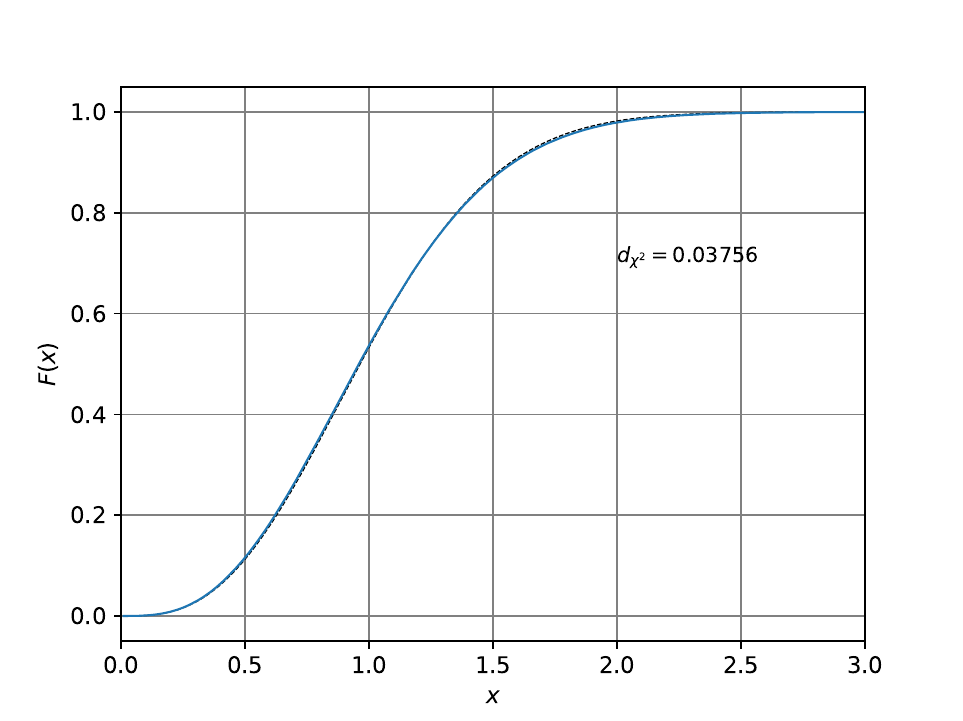}};
			\draw (-1.,2.9) node[black,fill=white,rounded corners=2pt,inner sep=1pt,opacity=1,text opacity=1]{$d_{\chi^2}=0.20144$};
		\end{tikzpicture}
			\caption{The level spacing distribution (left) and the cumulative density function (right), respectively, of the largest two eigenvalues of large random matrices generated via Monte Carlo simulations with $n=50,\,\nu=0$, $k=49$, with $n_{\rm conf.}=10^6$ configurations and bin size of about $0.2$. 
Because we compare to the spacing of a specific level $k$, here and in the following the number of level spacings is always equal to the number of configurations.
			The dashed curves are the bulk spacing distribution~\eref{eq:bulk-spacing} and its cumulative density $F(x)$.}\label{fig:soft-edge}
		\end{figure}

		Figure~\ref{fig:Comp_As-Bulk} highlights that for $k\geq3$ the deviation between the level spacing distributions at the hard edge (Eq.~\eref{lim-level-space}) and those in the bulk (Eq.~\eref{eq:bulk-spacing})  is comparable to the deviation between the latter and the Wigner surmise~\eref{Wignersurmise}. We have further quantified this result via a  $\chi^2$-test in Table~\ref{Tab:chi-sqrd-k}, where we have computed the $L^2$-distance between level spacing distributions~\eref{lim-level-space} and~\eref{eq:bulk-spacing} for various $k$ with Monte Carlo simulations. 
		
		 It can be seen, that for the transition from the hard edge to the bulk region of the spectrum, the deviation between the level spacing and the bulk spacing decreases, as it should, and increases again when reaching the upper edge at $k=n-1=49$ which is a soft edge and, thence, has to follow the Airy statistics.
		
		If we now look at the level spacing distribution at the soft edge (see Figure \ref{fig:soft-edge}) and the transition from the bulk to the latter, we find this transition to be equally fast. The differences obtained from the $\chi^2$-test between the numerical data and the bulk spacing are of the same order as for a spacing in the middle of the bulk, from the third spacing $k=3$ upwards and  
 from $k=47$ downwards,  
		as can also be seen in Table~\ref{Tab:chi-sqrd-k}.
		
		\begin{table}[h!]
			\centering
			\begin{tabular}{c||c|c|c|c|c|c|c|c|c}
				$k$				& 1		&	2		&	3		&	4		&	24		&	46		&	47		& 48		&	49	\\\hline
				$d_{\chi^2}$	&	0.08474	&	0.00927	& 0.00478	&	0.00413	&	0.00266	&	0.00333	&	0.00427	&	0.03756	&	0.20144
			\end{tabular}
			\caption{The $\chi^2$-test for the level spacing distributions between the $k$th and $(k+1)$st eigenvalue for $\nu=0$ at matrix dimension $n=50$, in comparison to the bulk spacing distribution~\eref{eq:bulk-spacing}. We have chosen a bin size of roughly $0.2$ and an ensemble size $n_{\rm conf.}=10^6$ to keep the statistical error below one percent. 
			}\label{Tab:chi-sqrd-k}
		\end{table}

	\subsection{Asymptotics to the Airy statistics ($\nu\gg1$)}\label{sec:asymp.nu}
		
		\begin{figure}[t!]
			\centerline{\includegraphics[width=0.49\textwidth]{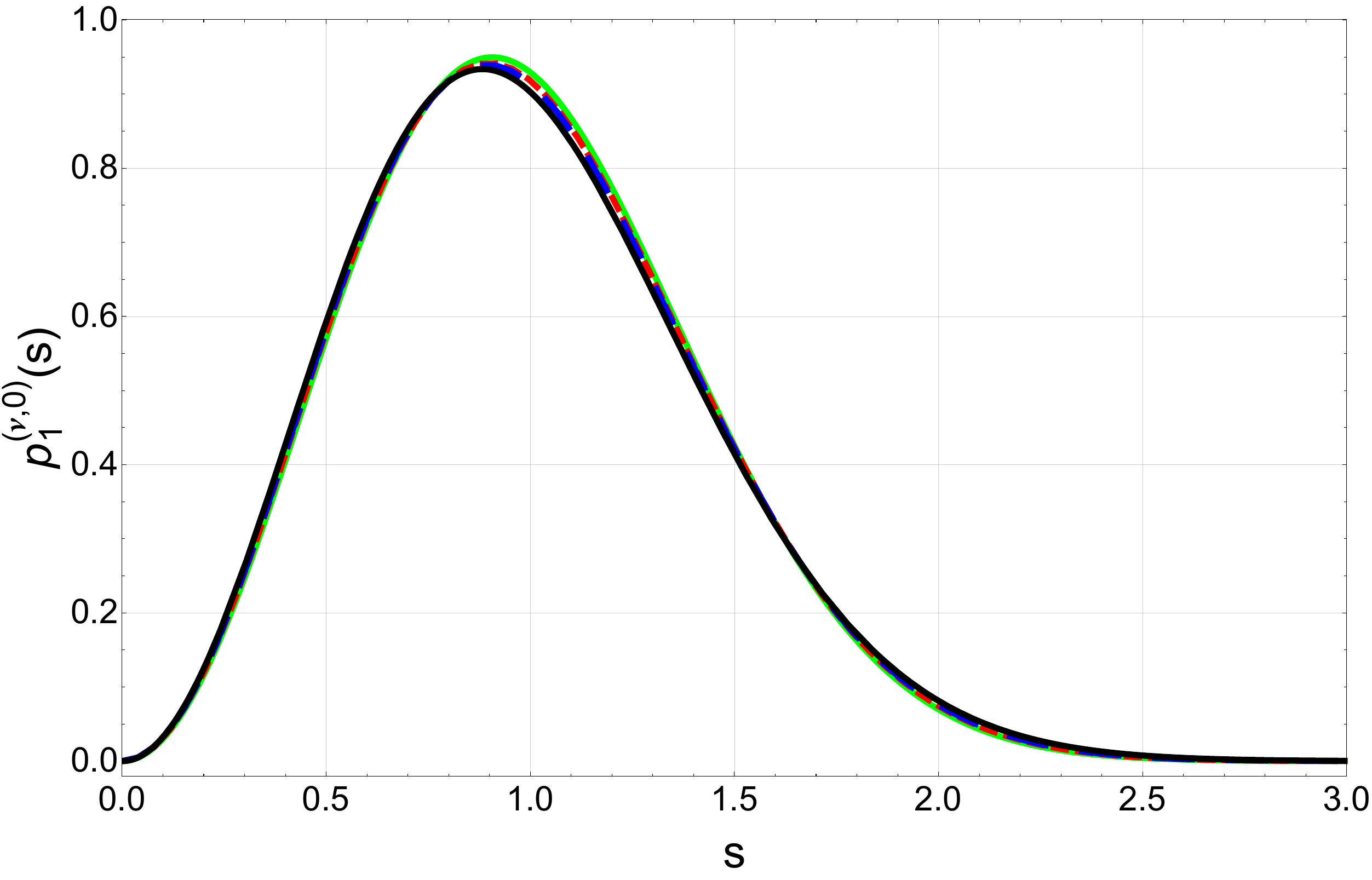}	\includegraphics[width=0.49\textwidth]{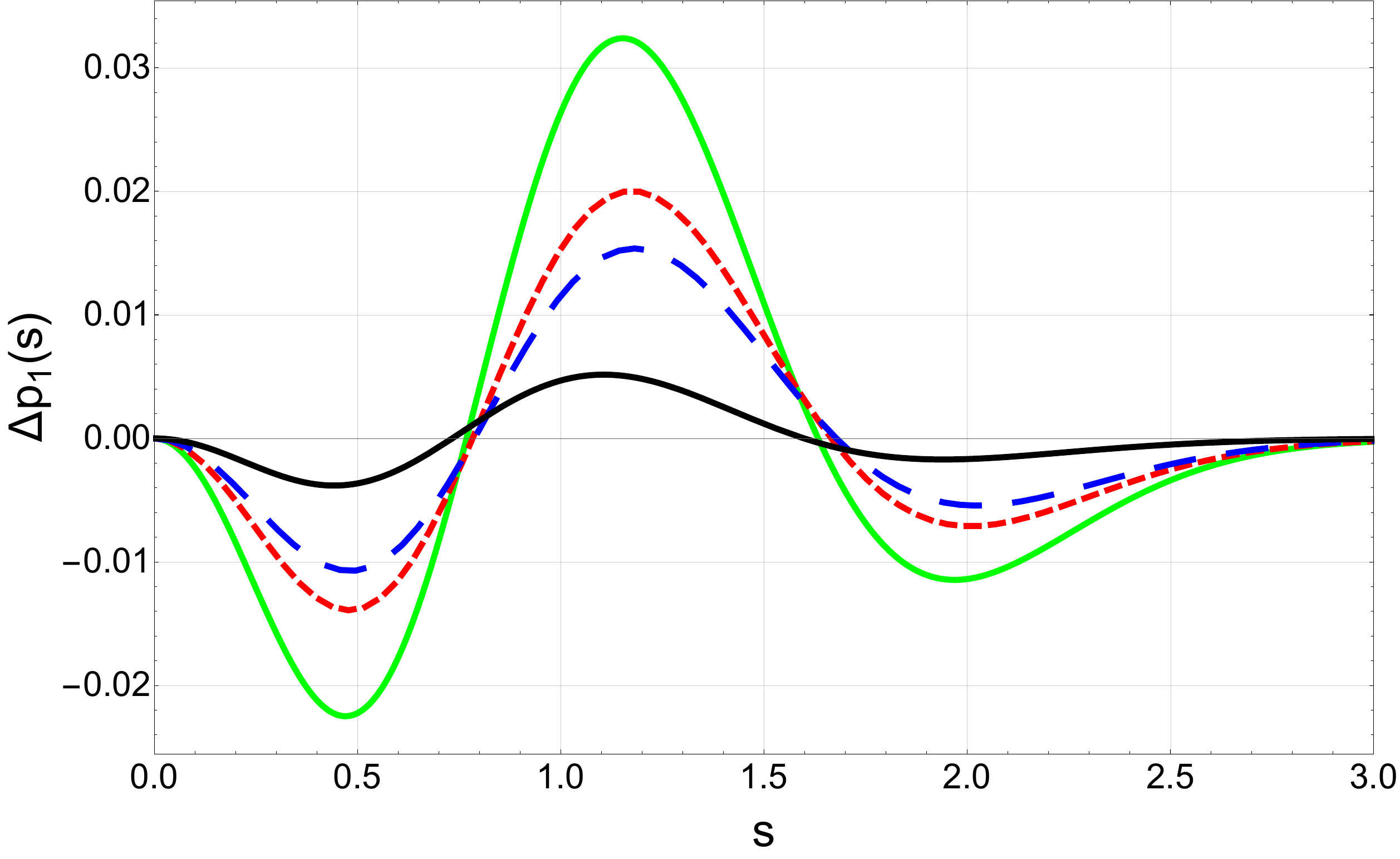}		}
			\centerline{\includegraphics[width=0.49\textwidth]{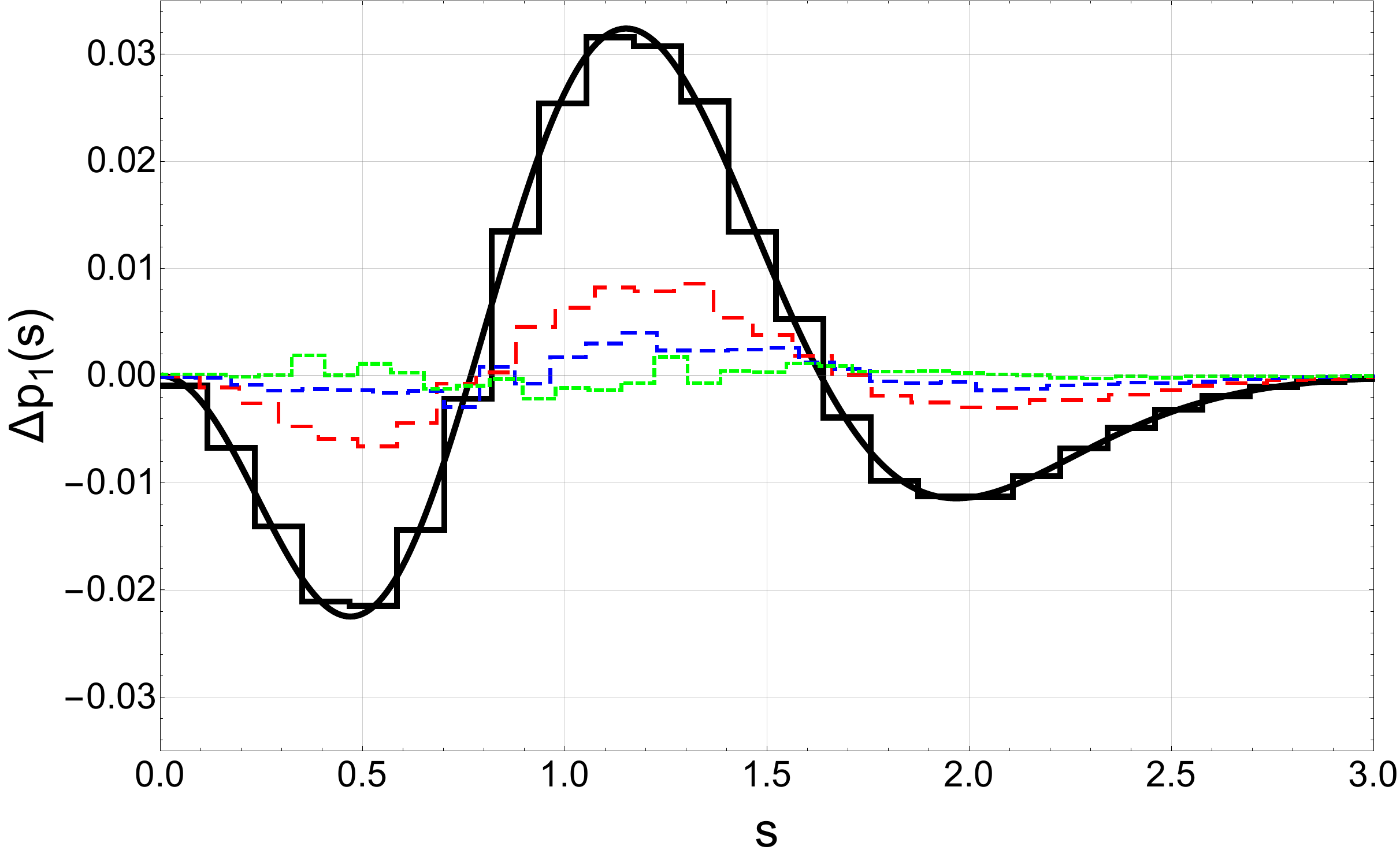}	\includegraphics[width=0.49\textwidth]{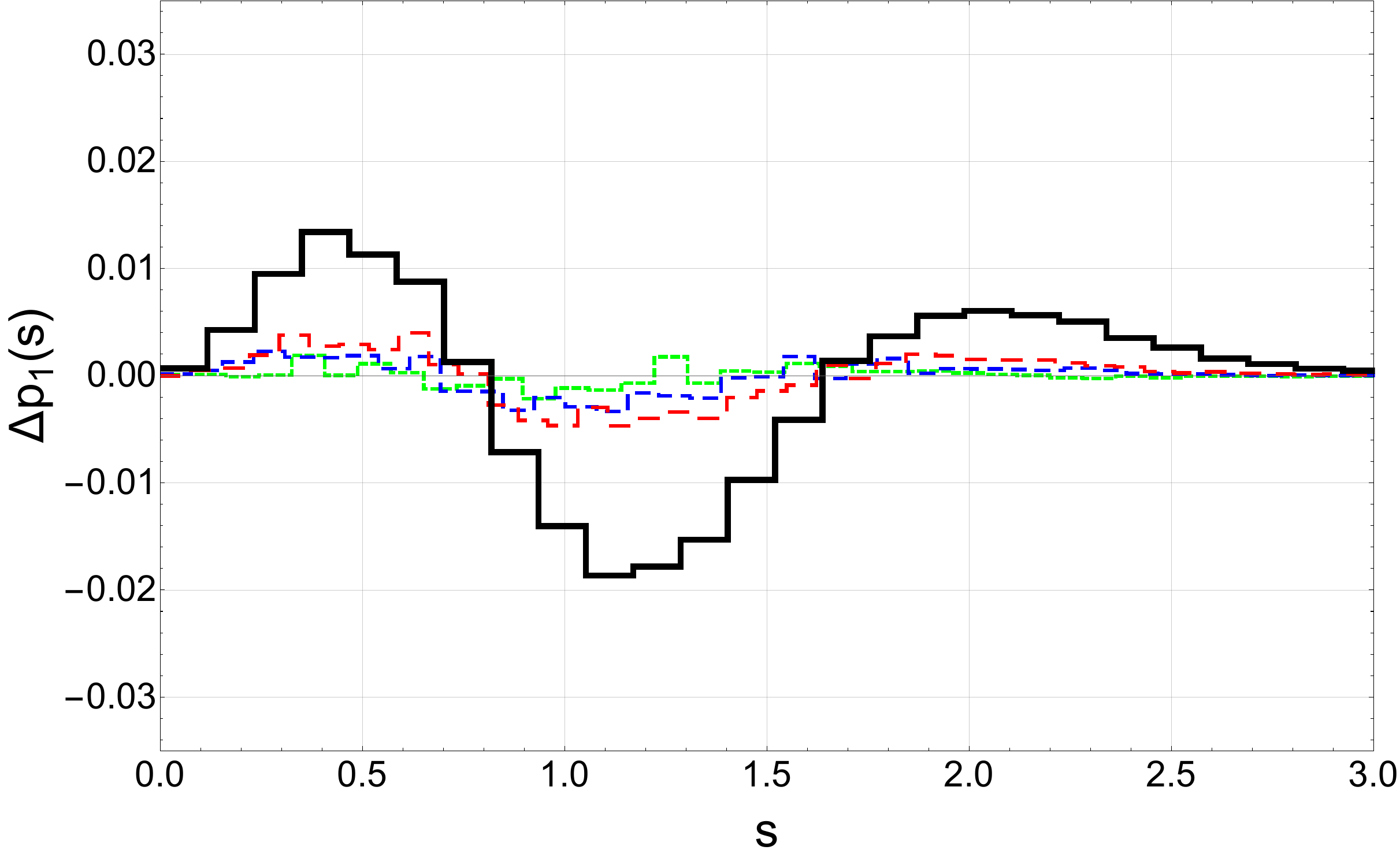}		}
			\caption{{\bf Top:} Level spacing distribution~\eref{lim-level-space} (left plot) and its difference to the bulk spacing distribution~\eref{eq:bulk-spacing} (right plot) for $N_{\rm f}=0$, $k=1$ and $\nu=0$ (green solid curves), $\nu=1$ (red finely dashed curves), and $\nu=2$ (blue coarsely dashed curves). The black curve in the left plot is the bulk spacing distribution~\eref{eq:bulk-spacing} and the one in the right plot is the difference of the Wigner surmise~\eref{Wignersurmise} to the  bulk spacing distribution as references.
			\newline {\bf Bottom:} Difference $\Delta p_1^{(\nu,0)}(s)=\bar  p_1^{(\nu,0)}(s)-p_{\rm bulk}(s)$ of the unfolding~\eref{unfolded-level-density} of the level spacing distribution $p_1^{(\nu,0)}$ (see~\eref{lim-level-space}), and the bulk spacing distribution $p_{\rm bulk}(s)$ (see~\eref{eq:bulk-spacing}) for  the two smallest eigenvalues at $N_{\rm f}=0$ and various $\nu$. All histograms are generated by Monte Carlo simulations with $10^7$ configurations, a matrix dimension $n=100$ and a bin size of roughly $0.15$.  The values of $\nu$ are as follows: {\bf Left:} $\nu=0$ (black solid histogram, solid curve is the analytical result~\eref{lim-level-space}), $\nu=5$ (roughly dashed red), $\nu=10$ (dashed blue), $\nu=15$ (finely dashed green); {\bf Right:} $\nu=15$ (finely dashed green), $\nu=20$ (dashed blue), $\nu=25$ (roughly dashed red). For comparison, the unfolded level spacing distribution between the largest two eigenvalues $\Delta p_{49}^{(0,0)}(s)$ (solid black) is added, corresponding to $\lim_{\nu\to\infty}\Delta p_1^{(\nu,0)}(s)$). We have plotted $\nu=15$ in both plots as a reference value 
for the transition.			
						}\label{fig:nu-depend}
		\end{figure}
	
Another limit is the large-$\nu$ limit which should approach the level spacing distributions at the soft edge.  Actually this also holds when the number of flavours $N_{\rm f}$ is sent to infinity, while the corresponding masses are  of the order of $1/\sqrt{n}$, as we have exploited this scaling in our computations. In general, we can understand the effect of non-vanishing flavours and topology as follows. Both push the eigenvalues further away from the origin, making the effect of the hard wall less relevant.
		
		\begin{figure}[t!]
			\centering
			\includegraphics[width=0.45\textwidth]{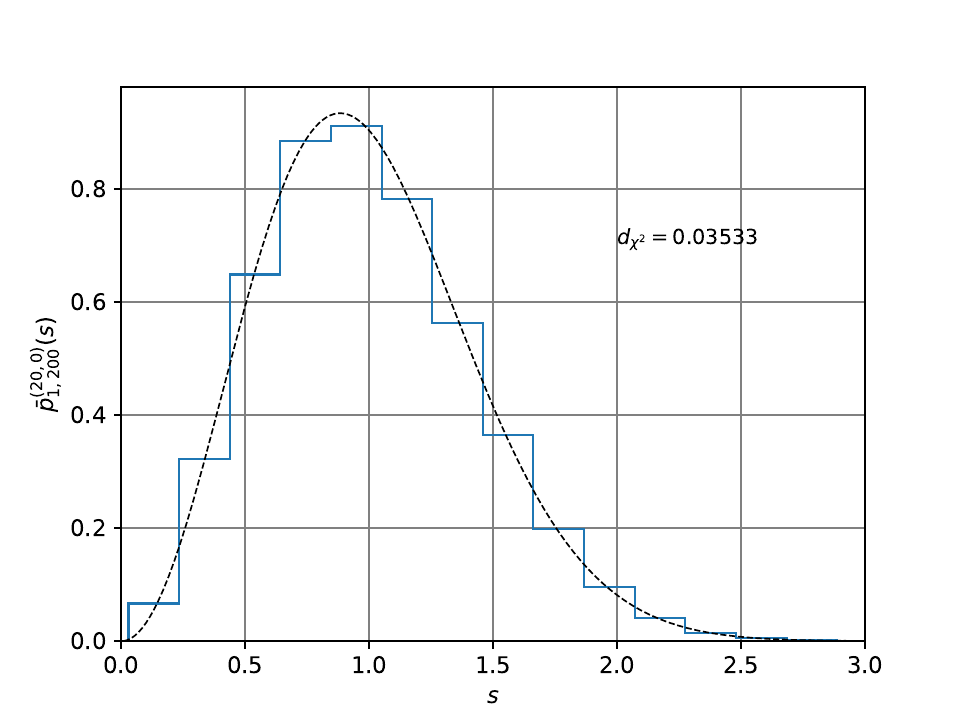}		
			\includegraphics[width=0.45\textwidth]{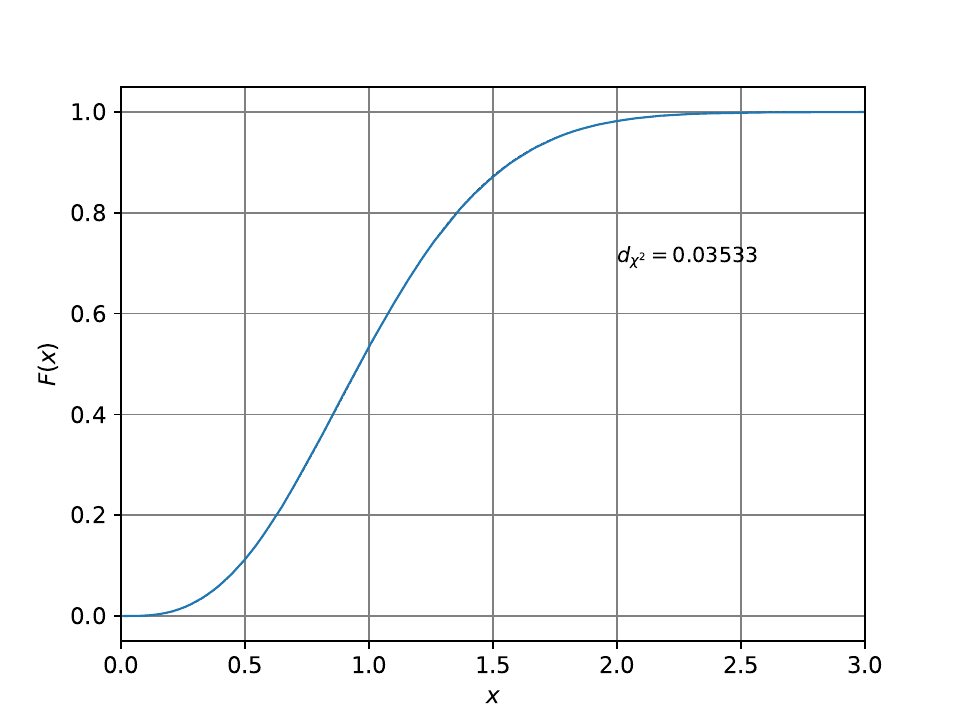}	
			\includegraphics[width=0.45\textwidth]{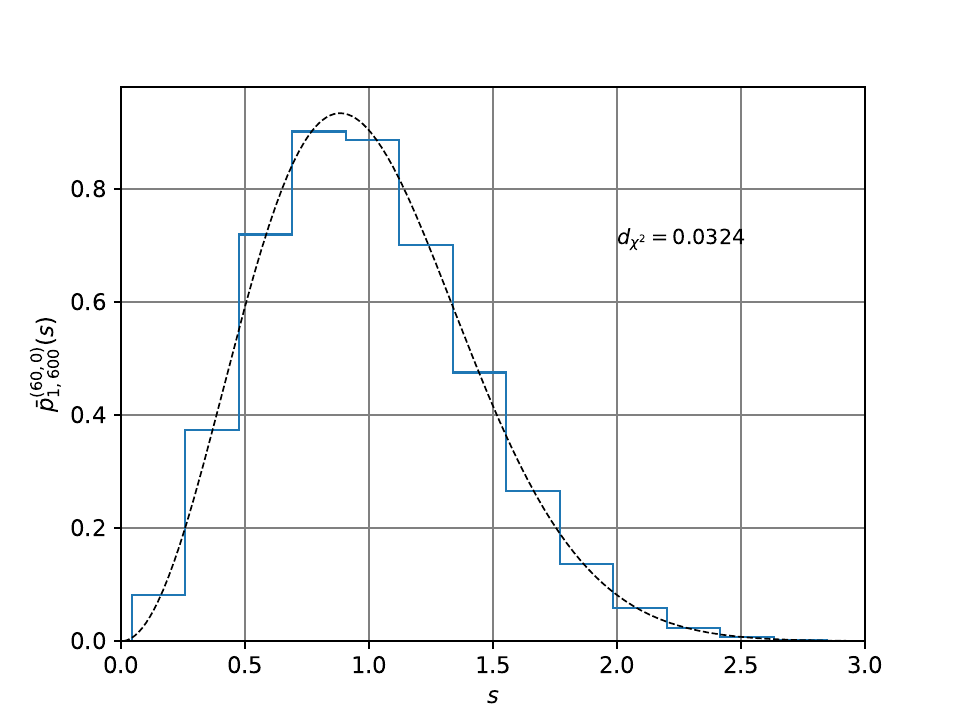}		
			\includegraphics[width=0.45\textwidth]{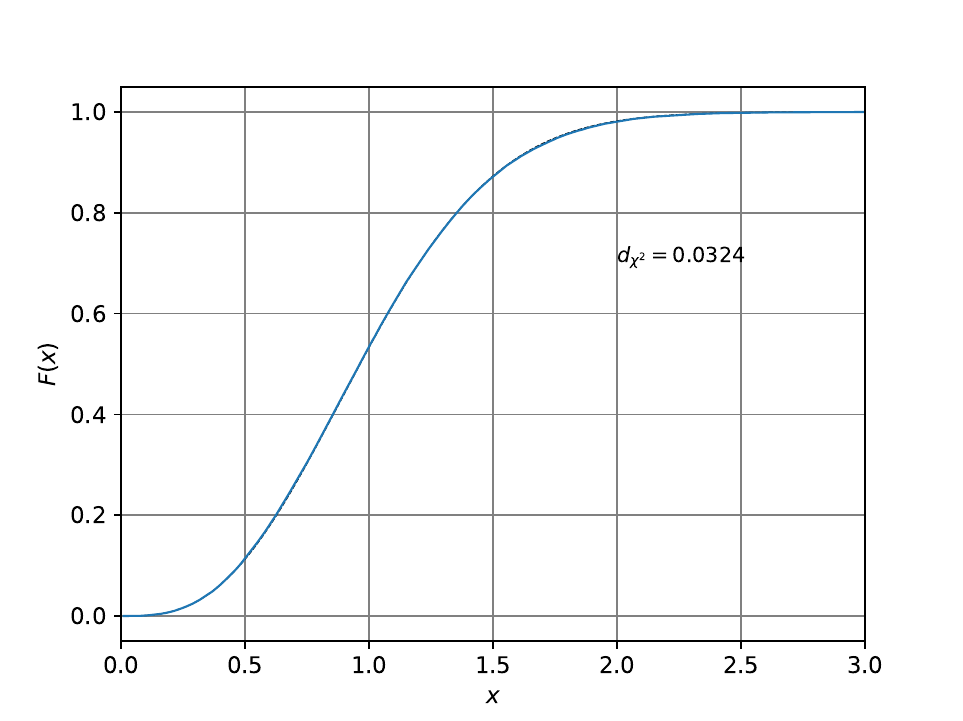}	
			\includegraphics[width=0.45\textwidth]{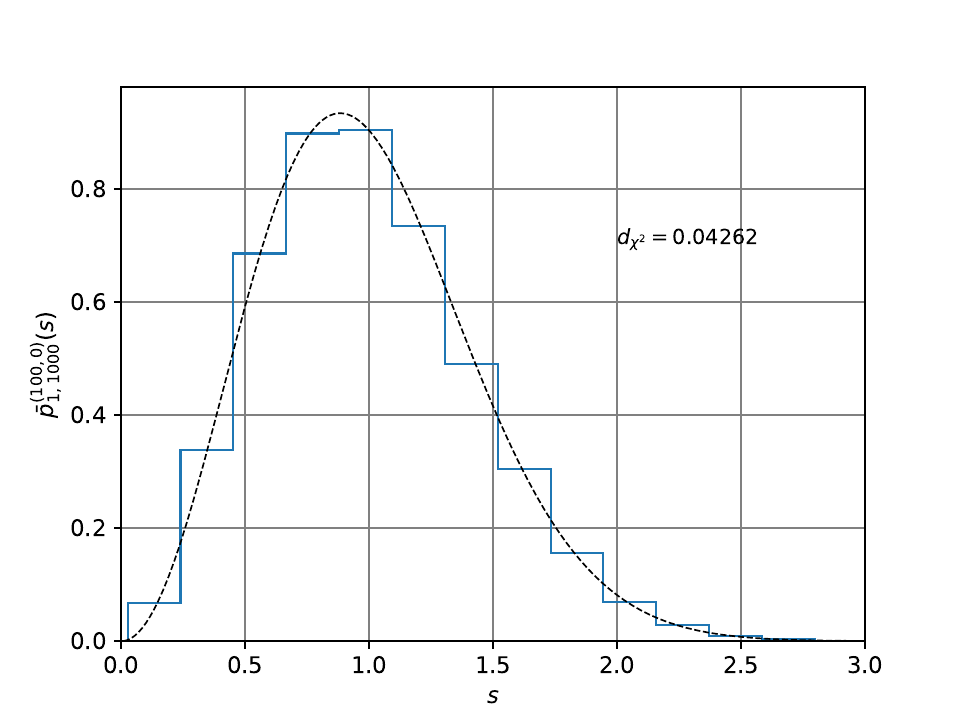}
			\includegraphics[width=0.45\textwidth]{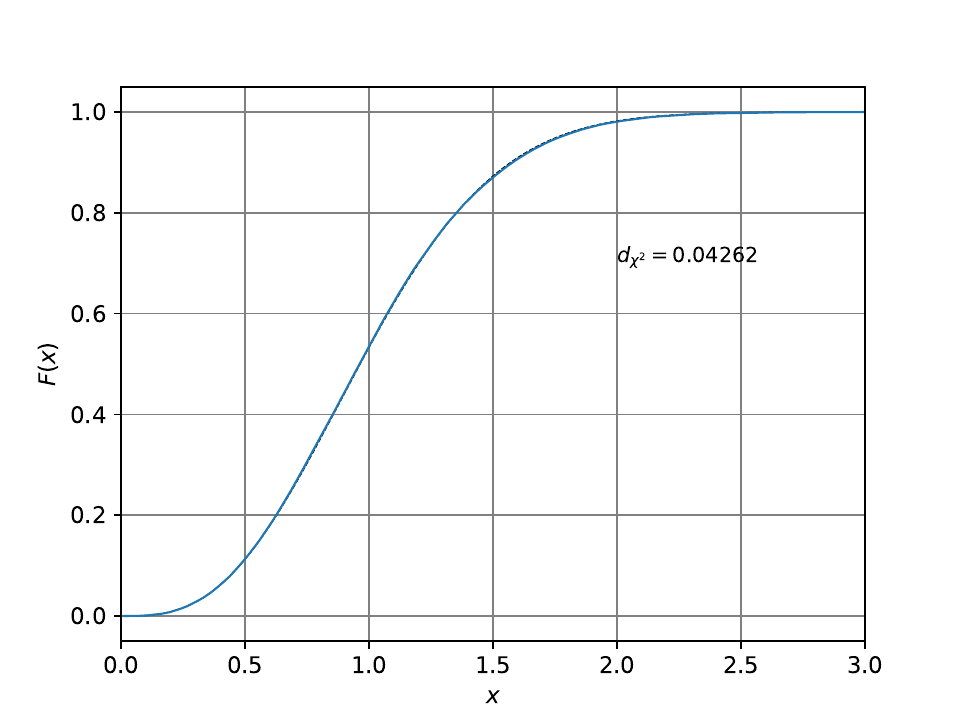}	
			\caption{The level spacing distributions (left column) and cumulative densities (right column) of the smallest two eigenvalues ($k=1$) generated via Monte Carlo simulations for $N=200,\,\nu=4$ (top), $N=600,\,\nu=12$ (middle) and $N=1000,\,\nu=20$ (bottom). The number of configurations is  $n_{\rm conf.}= 10^5$, as the matrix size is already very big, and the bin size varies about $0.15-0.2$ due to different rescaling from the unfolding for each parameter set. In the insets we have given the $L^2$ distance to the bulk spacing distribution~\eref{eq:bulk-spacing} and also its cumulative density,  which are drawn as dashed curves. 
}\label{fig:Large_nu-Bulk}
		\end{figure}
			
In the top plots of Figure~\ref{fig:nu-depend} we have numerically evaluated~\eref{lim-level-space} for $k=1$, $N_{\rm f}=0$ and $\nu=0,1,2$. Hence, it is the level spacing between the two smallest eigenvalues for a different number of zero eigenvalues. As mentioned before, we are unable to go much beyond this setting. Thence, we have generated Monte Carlo simulations with matrix size $n=100$ for larger values of $\nu$. Those are fairly good approximations of the asymptotic result, as the  rate of convergence for the unfolded spacing distribution is $1/n^2$ at the hard edge, cf.~\cite{ForresterTrinh}. As all unfolded level spacing distributions are very close together, compare the upper left plot in Figure~\ref{fig:nu-depend}, we have studied the differences to the bulk spacing distribution $p_{\rm bulk}(s)$, see~\eref{eq:bulk-spacing}, as a function of $\nu$. For the histograms from the Monte Carlo simulations one needs to integrate $p_{\rm bulk}(s)$ over the length of the bins, so that one can take such a difference. Thus, the general procedure has been as follows: (1) creating the histogram of the level spacing distribution $p_1^{(\nu,0)}(s)$ of the simulation; (2) unfold it via the formula~\eref{unfolded-level-density}; (3) use the unfolded bins to create a similar histogram of~\eref{eq:bulk-spacing}; (4) finally take the difference $\Delta p_1^{(\nu,0)}(s)=\bar  p_1^{(\nu,0)}(s)-p_{\rm bulk}(s)$.

One can readily see in Figure~\ref{fig:nu-depend} (bottom plots), that the distance between the level spacing distribution of the two smallest eigenvalues and the one in the bulk first decreases with increasing $\nu$ (left plot). However, at about $\nu=15$ it again increases  as the distribution slowly approaches the level spacing distribution of the soft edge, which we have added for comparison, see lower right plot in Figure~\ref{fig:nu-depend}. 
The spacing distribution of the two largest eigenvalues is also Monte Carlo simulated for the matrix dimension $n=100$. However, one needs to be careful with this result, as the finite size error is quite big at the soft-edge. In~\cite{ForresterTrinhsoft} it was found that the optimal rate of convergence is $n^{-2/3}$. Thus, in the present case the systematic error is about $20\%$ and, therefore, much larger than the deviations to the bulk level spacing distribution~\eref{eq:bulk-spacing} we are comparing with. To have it of the same order one needs to generate matrices of the size $n=10^5$ which is out of reach for our modest computing power.
	
		As already discussed in Sec.~\ref{sec:intro}, it is well known~\cite{CK}, that for large $\nu$ the Bessel-kernel converges to the Airy-kernel, as the hard edge of the Marchenko-Pastur law deforms into a second soft edge. To get a statistical quantification of this effect, we have performed another Monte Carlo simulation of the random matrix ensemble with $\nu\sim n$ for increasing $\nu$, and calculated their spacings, differences to the bulk and empirical densities to observe this effect (see Figure~\ref{fig:Large_nu-Bulk}). We further quantify this transformation again in terms of the $\chi^2$-test comparing the statistics to the bulk spacing in Table~\ref{Tab:chi-sqrd-nu}. In the latter, it becomes apparent, that the deviation to the bulk spacing increases rapidly for decreasing values of $q= n / (n + \nu)$, which shows a rapid deformation of the hard edge into a soft edge. This has been already seen in Figure~\ref{fig:Bessel-Airy-trans} for the microscopic level density, when $\nu=20$.
	
		\begin{table}[h!]
			\centering
			\begin{tabular}{c||c|c|c}
				$\ulbackslash{q}{n}$&	200		&	600		&	1000	\\\hline\hline
						0.98		&	0.01286	&	0.00283	&	0.00527	\\
						0.91		&	0.03533	&	0.0324	&	0.04262	\\
						0.83		&	0.01056	&			&
			\end{tabular}
			\caption{The $\chi^2$-test for the level spacing of the smallest two eigenvalues for a few decreasing values of $q = n / (n + \nu)$ and increasing matrix dimension. The number of configurations is varying between $n_{\rm conf.}=10^5$ and $n_{\rm conf.}=10^6$, where for the lowest $q$ we only went up to $n=200$.
			}\label{Tab:chi-sqrd-nu}
		\end{table}
	
		If we compare this to spacings at the soft edge for $\nu=0$ (Table~\ref{Tab:chi-sqrd-k}), we find distances of the same order from the $\chi^2$-test, which shows that for larger $\nu$ the hard edge transforms into a soft edge.
		
		In Figure~\ref{fig:Large_nu-Bulk} as well as Figure~\ref{fig:nu-depend}, it becomes apparent, that 
when moving from the hard to the soft edge 		
		the peak of the spacing shifts from the right to the left of the maximum of the bulk spacing distribution. This is expected as a stiffer level spacing, as it is the case for the hard edge, should have a maximum closer to $1$. In contrast,  for a softer level spacing, like for the Airy statistics, the eigenvalues  have more freedom to move, which shows in a maximum further away from $1$. Since this effect is barely noticeable, the value of the $\chi^2$-test is additionally given in Figure~\ref{fig:Large_nu-Bulk}. We omitted to show the difference to the bulk spacing distribution as the number of generated configurations is smaller here. This is because the matrix size in the second setting of Monte Carlo simulations has been increased already. This results in a statistical error of one percent, which strongly overshadows the deviation from the bulk statistics.
			\begin{figure}[t!]
		\centering
		\includegraphics[width=0.49\textwidth]{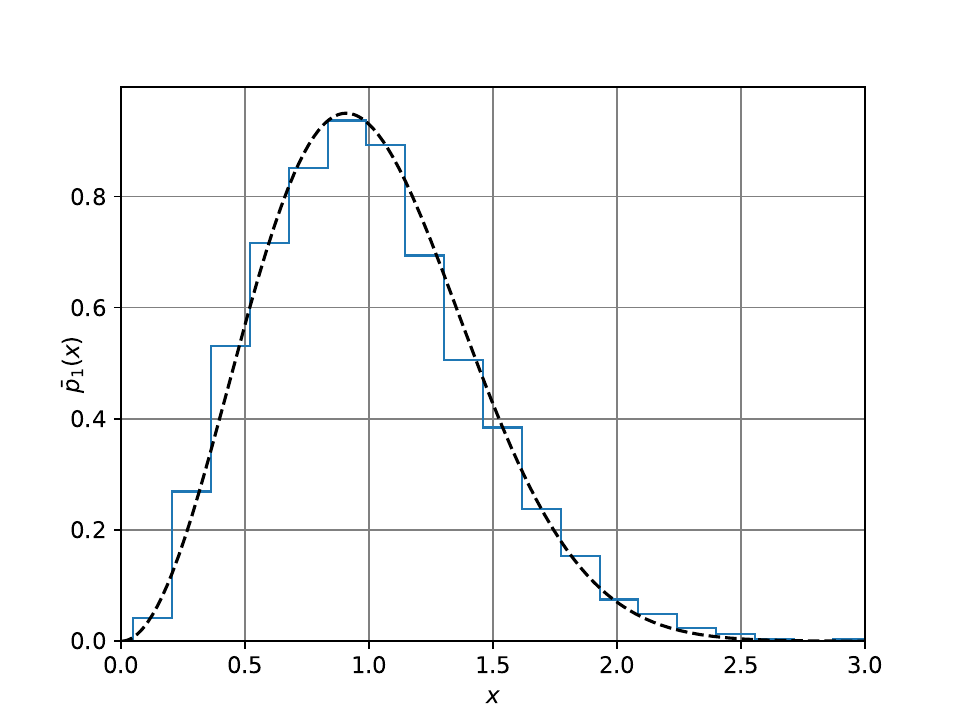}
		\includegraphics[width=0.49\textwidth]{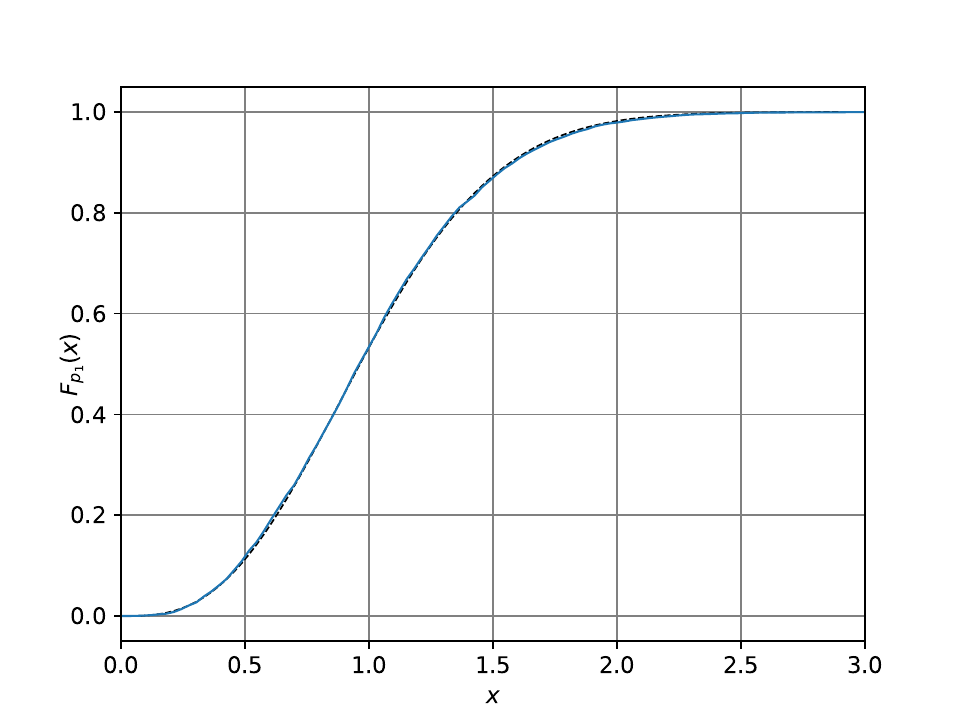}
		\caption{{\bf Left:} The level spacing distribution  of the smallest two eigenvalues of the overlap Dirac operator $k=1$  (blue histogram) with $N_f=2$ and $\nu=0$ averaged over six different degenerate masses, with a total number of configurations of $n_{\rm conf.}=6046$. It compared to the spacing distribution between the two smallest eigenvalues of the $\chi$GUE in the large-$n$ limit for $N_{\rm f}=0$ quenched with $\nu=0$ (dashed curve). {\bf Right:} The cumulative densities of the two distributions, respectively.
}\label{fig:comp_k_2_QCD1}
	\end{figure}

\section{Comparison to Data from Lattice QCD}\label{subsec:LQCD_data}

Finally, we would like to compare with edge statistics from data, notably at the 
hard edge where we have derived analytical predictions. 
For that reason we consider empirical data from lattice QCD Dirac operators.

It is known~\cite{TJ} that strongly interacting quantum field theories in the deepest infrared limit, the $\varepsilon$-regime, agree with the statistics obtained from 
random matrix theory. In QCD, as it appears in the standard model, namely with the gauge group ${\rm SU}(3)$ in the fundamental representation and a four dimensional Euclidean space-time,  the corresponding random matrix theory is the $\chi$GUE. For our purposes this means that the $\chi$GUE provides a description for the statistics of the smallest eigenvalues and their spacings for the QCD Dirac operator. This carries over to most lattice discretisations of this theory.

We gathered data from the JLQCD collaboration described in \cite{JLQCD}. This data is obtained from numerical simulations on a Euclidean $16^3\times 32$ space-time lattice at $\beta=2.30$, with a lattice spacing $a\sim 0.12$fm, via two different algorithms, namely domain wall and overlap fermions. From those two algorithms the $50$ smallest eigenvalues of the overlap Dirac operator with a total number of configurations of about $n_{\rm conf.}=1000$ 
were computed for different values of Fermion masses, where we will focus on the data with $N_f=2$, see top part of table 1 in \cite{JLQCD}. 

Although 
we have derived detailed expressions for the spacing distribution as a function of the number of flavours $N_f$ and rescaled quark masses $\mu_f$ \eref{lim-level-space}, we have also seen that the difference between quenched and unquenched prediction is very small, see Fig. \ref{fig:Comp_As-Bulk}. Furthermore, it is evident from the seminal paper \cite{BGS}, that even for about 2000 spacings only a rather coarse comparison to the (bulk) spacing can be made. 
To further increase the statistics from \cite{JLQCD} which is already exceptionally high for a given set of parameters, we have therefore decided to average over all given mass configurations 
resulting in $n_{\rm conf.}=6046$. 
The mass range in units of lattice spacing of the data \cite{JLQCD} is $m_{ud}=0.015,0.025,0.035,0.05,0.07,0.1$, which maps to a range of about $\mu_f=1-50$.  
Thus not all of these masses are large enough to apply the quenched approximation $N_f=0$ to the $\chi$GUE in \eref{lim-level-space}.

	In Figure~\ref{fig:comp_k_2_QCD1}, we find a quite large deviation to the quenched spacing distribution \eref{lim-level-space00k2} at $\nu=0$ at the hard edge, of up to $8\%$. The $L^2$-distance given by the $\chi^2$-test is given by $d_{\chi^2} = 0.44616$.
	Here, one has to keep in mind, that this is to a certain extend due to the number of configurations of lattice simulations available - despite averaging over different masses. The statistical error exceeds by one order of magnitude the difference between the spacing distributions  in the bulk, at the hard and soft edge. Therefore, we can conclude that the level spacing distribution is not at all a good measure in lattice QCD to discern hard edge from bulk statistics. 
This is both because the effect of unquenching is so small and the number of configurations to 
detect
this is exceedingly high.	

Only deviations due to a mobility edge 
are visible as found in~\cite{Poul,Kim,NGTKP,KovacsPittler}. 
Here, the mobility edge denotes the boundary between 
localised states, associated with Poisson statistics, and delocalised states related to random matrix statistics. In QCD it is reached by increasing the temperature towards the chiral phase transition, as studied in ~\cite{Poul,Kim,NGTKP,KovacsPittler}. 
Close to the mobility edge, 
the global symmetries of the states are a mixture with Poisson statistics, which diminishes the level repulsion drastically and yields deviations from the bulk level spacing above the percent threshold, that can be easily discerned by the bare eye.

\section{Conclusion}\label{sec:conclusio}

In the application of random matrices we have learned to appreciate the predictive power for spectral correlations, depending on the location within the spectrum. One example is the microscopic Bessel-density close to the origin, that is relevant when chiral symmetry is important, e.g. in comparison to data from lattice QCD. It depends heavily on the number of zero-modes $\nu$, that characterise different topological sectors of the theory with broken chiral symmetry, and on the number of light flavours $N_{\rm f}$  when the quark masses are sufficiently small. A second example is the Tracy-Widom distribution for the largest eigenvalue, that is relevant in the vicinity of a soft spectral edge. It describes successfully the fluctuations of growth processes for example.

In this work we have learned, that in contrast to that, the spacing distribution between consecutive smallest or largest eigenvalues is a rather inefficient measure to quantify the specific properties of these spectral regions. We could provide compact expressions for the spacing to the $k$th smallest eigenvalue, that explicitly depend on $k$, the number of zero-modes $\nu$ and number of flavours $N_{\rm f}$. However, once evaluated these turned out to be very close to the bulk spacing for $k=1$ and $\nu=0=N_{\rm f}$ already, converging very rapidly as close to the exact spacing as the Wigner surmise, almost independently of $\nu$ and $N_{\rm f}$.
We have demonstrated this lack of distinction directly upon comparing with QCD lattice data, that are known to  follow the predictions of the $\chi$GUE for the microscopic density and individual eigenvalues distributions in the $\varepsilon$-regime.

We expect that the very same feature persists when considering chiral ensembles of random matrices with orthogonal or symplectic symmetry.  The spacing distribution in these two ensembles is different from the GUE in all parts of the spectrum.  When comparing the respective spacing distributions at the hard and soft edge to the bulk, the result is probably equally close as found here in the unitary symmetry class.  Substantial deviations from the bulk spacing at the spectral edges can only be expected when considering ensembles with heavy tails or other substantial deformations.

\section*{Acknowledgments}

This work was funded by the Deutsche Forschungsgemeinschaft (DFG, German Research Foundation) – SFB 1283/2 2021 – 317210226 "Taming uncertainty and profiting from randomness and low regularity in analysis, stochastics and their applications" (GA) and by the Australian Research Council (ARC) via the grant DP210102887 (MK).  Moreover, we would like to thank Hidenori Fukaya for providing us with the lattice data on behalf of the JLQCD collaboration.

\appendix

\section{The Spacing Distribution in the Bulk}\label{spaceDH}

There are several alternative formulations of the exact spacing distribution for the GUE in the bulk available. For instance in Mehta's book \cite{Mehta}, in chapter 6 an infinite product expansion in terms of Fredholm determinant eigenvalues is offered. The latter are given by integrals of spheroidal functions. 

The bulk spacing 
for the GUE
used in the present work is taken from~\cite{DH}. It is approximated by
a highly accurate Pad\'e expansion~\cite{Pade}.  Such an expansion is given in terms of two rational functions, that takes into account the correct small- and large-$s$ behaviour. In the present case this is given by
\begin{eqnarray}
		\fl p_{\rm bulk}(s) = \frac{\pi^4}{16}\frac{f(s)}{g(s)} \left(s^2 - \frac{2}{\pi^2} + \frac{5}{\pi^4 s^2}\right) 
						\left(\frac{\pi  s}{2}\right)^{-\frac{1}{4}}\exp{\left[\frac{\log{(2)}}{12} + 3 \left(\frac{1}{12} - \log{(1.2824271291)}\right) - \frac{\pi^2 s^2}{8}\right]}.
\label{eq:bulk-spacing}
\end{eqnarray}
The two functions $f(s)$ and $g(s)$ are polynomials in $s^{1/4}$ and when choosing them of maximal degree $44$ they are given by~\cite{DH}
\begin{eqnarray}
		f(s) &=& 28915.18572129 s^{\frac{17}{4}} + 1086.6573434 s^{\frac{20}{4}} + 45351.83787784 s^{\frac{21}{4}} + 1772.4275928 s^{\frac{24}{4}}\nonumber \\
			&&+ 30526.8898691 s^{\frac{25}{4}} + 1242.5658114 s^{\frac{28}{4}} + 10452.3148571 s^{\frac{29}{4}} + 446.60040377 s^{\frac{32}{4}}\nonumber \\
			&&+ 3018.6826326 s^{\frac{33}{4}} + 125.982683 s^{\frac{36}{4}} + 346.045278 s^{\frac{37}{4}} + 16.21088391 s^{\frac{40}{4}}\nonumber \\
			&&+ 31.04144855 s^{\frac{41}{4}} + s^{\frac{44}{4}},\\
		g(s) &=& 1582.4460446 + 59.4696721 s^{\frac{3}{4}} + 2481.97736506 s^{\frac{64}{4}} + 96.9999314 s^{\frac{7}{4}}\nonumber \\
			&&- 4446.44557011 s^{\frac{8}{4}} - 161.88357063 s^{\frac{11}{4}} - 9022.29390885 s^{\frac{12}{4}} - 350.52128316 s^{\frac{15}{4}}\nonumber \\
			&&+ 23929.7407678 s^{\frac{16}{4}} + 879.81693061 s^{\frac{19}{4}} + 45324.33326465 s^{\frac{20}{4}} + 1763.26689471 s^{\frac{23}{4}}\nonumber \\
			&&+ 34256.40313293 s^{\frac{24}{4}} + 1384.64210377 s^{\frac{27}{4}} + 14982.65463299 s^{\frac{28}{4}} + 625.59278179 s^{\frac{31}{4}}\nonumber \\
			&&+ 4081.38643920 s^{\frac{32}{4}} + 175.06649190 s^{\frac{35}{4}} + 682.7954698 s^{\frac{36}{4}} + 29.56032569 s^{\frac{39}{4}}\nonumber \\
			&&+ 57.1857440 s^{\frac{40}{4}} + 2.36447174 s^{\frac{43}{4}} + s^{\frac{44}{4}}.
\end{eqnarray}
We would like to underline that this expansion is of the exact bulk level spacing distribution. Hence, the approximation is not exactly normalised to norm and first moment of unity. Dietz and Haake~\cite{DH} mention that the expansion, up to the terms we have shown here, yields a deviation from the norm and first moment that lie below $10^{-8}$ and $10^{-11}$, respectively. This can be also seen as a measure for the deviation from the exact spacing distribution.

A similar expansion for the bulk spacing in the Gaussian orthogonal and symplectic ensemble can be found in~\cite{DH} as well.


\end{document}